\documentclass[reqno,11pt]{amsart}\usepackage[]{graphicx}\usepackage[]{color}
\makeatletter
\def\maxwidth{ %
  \ifdim\Gin@nat@width>\linewidth
    \linewidth
  \else
    \Gin@nat@width
  \fi
}
\makeatother

\definecolor{fgcolor}{rgb}{0.345, 0.345, 0.345}

\usepackage{framed}
\makeatletter
 {\par\unskip\endMakeFramed%
 \at@end@of@kframe}
\makeatother

\definecolor{shadecolor}{rgb}{.97, .97, .97}
\definecolor{messagecolor}{rgb}{0, 0, 0}
\definecolor{warningcolor}{rgb}{1, 0, 1}
\definecolor{errorcolor}{rgb}{1, 0, 0}

\usepackage{alltt}

\usepackage{accents,setspace,graphicx,srcltx,enumitem}
\usepackage[round]{natbib}
\usepackage{subfigure}
\usepackage{bbm}
\usepackage{multirow}
\usepackage{amsmath}
\usepackage{amssymb}
\usepackage{amsfonts}
\usepackage{amsaddr}
\usepackage[english]{babel}
 
\usepackage{MnSymbol}  
\usepackage{hyperref}

\newcommand{\field}[1]{\mathbb{#1}}
\newcommand{\R}{\field{R}}

\newtheorem{lemma}{Lemma}[section]
\newtheorem{theorem}{Theorem}[section]
\newtheorem{proposition}{Proposition}[section]
\newtheorem{assumption}{Assumption}[section]

  \newtheorem{assumpB}{Assumption}[section]
  \newcounter{assB1}
  \newcounter{assB2}
  \theoremstyle{remark}
  \newtheorem{remark}[]{Remark}
  \newtheorem{example}[]{Example}

     \allowdisplaybreaks
   
     \newcommand{\SM}{\mathbf{S}}
     
     \newcommand{\PM}{\mathbf{P}}

\title{Semiparametric Estimation of Correlated Random Coefficient Models without Instrumental Variables}

\date{\today}
\author{Samuele Centorrino$^\ast$}
\address[S.\ Centorrino]{$(^\ast)$ Corresponding Author.\\
Economics Department, State University of New York at Stony Brook, USA.\\\textit{Email address}: \texttt{\textup{samuele.centorrino@stonybrook.edu}}.}

\author{Aman Ullah}
\address[A.\ Ullah]{Department of Economics, University of California Riverside, USA.\\\textit{Email address}: \texttt{\textup{aman.ullah@ucr.edu}}.}

\author{Jing Xue}
\address[J.\ Xue]{School of Economics, Dongbei University of Finance and Economics, China.\\\textit{Email address}: \texttt{\textup{xuejingdufe@163.com}}.}

\thanks{The authors would like to thank Juan Pantano and Peter Robinson for useful conversations, comments and remarks. The authors declare no conflict of interest.}

\IfFileExists{upquote.sty}{\usepackage{upquote}}{}
\begin{document}

\begin{abstract}
We study a linear random coefficient model where slope parameters may be correlated with some continuous covariates. Such a model specification may occur in empirical research, for instance, when quantifying the effect of a continuous treatment observed at two time periods. We show one can carry identification and estimation without instruments. We propose a semiparametric estimator of average partial effects and of average treatment effects on the treated. We showcase the small sample properties of our estimator in an extensive simulation study. Among other things, we reveal that it compares favorably with a control function estimator. We conclude with an application to the effect of malaria eradication on economic development in Colombia.\\


\end{abstract}

\maketitle

\doublespacing

\newpage 

\section{Introduction}

Linearity of a causal relationship between a dependent variable and a set of regressors is one of the pillars of empirical work in economics. Linear models are usually written under the assumption that individual agents respond homogeneously to a change in regressors, and cannot capture heterogeneity which seems to be a pervasive feature of many economic models. The linear random coefficient model allows a researcher to easily incorporate heterogeneous responses while maintaining the simplicity of the linear specification. 

Consider the following random coefficient model

\begin{equation}\label{eq:mainmodel}
Y_i = \alpha_i + X_i\beta_{i},
\end{equation}

where $X_i \in \R^p$ is a row vector of continuous regressors, and $\lbrace \alpha_i, \beta_{i} \rbrace$ is a $(p+1) \times 1$ column vector of random coefficients. In particular, we can write the vector of random coefficients as follows
\begin{align*}
\alpha_i =& \alpha + \eta_i\\
\beta_i =& \beta + \varepsilon_i, 
\end{align*}
where $\eta_i$ is a zero mean scalar random error term; and $\varepsilon_i$ is a $p\times 1$ column vector of zero mean random errors. We assume that $E \left[ \alpha_i\right] = \alpha < \infty$, and $E \left[ \beta_i\right] = \beta < \infty$. Empirical researchers are usually interested in the Average Partial Effects (APE), $\beta_j$, for $j = 1,\dots,p$, although the (conditional or unconditional) distribution of the random coefficients may also be a target. 

Whenever the random coefficients are independent of the vector of regressors, $X_i$, ordinary least squares yield consistent estimators of the APE. However, in many economic models, the individual response to a change in a regressor may depend on the regressor itself. \citet{wooldridge1997} and \citet{heckman1998} discuss this issue in a model of returns to schooling, where the marginal effect of schooling on wage may depend on the unobserved level of ability, and schooling and ability are correlated. In the latter case, a simple OLS estimator would be inconsistent. 

Most of the existing literature thus relies on an instrumental variable restriction for identification and estimation. Upon additional restrictions on the model and upon the validity of such an instrumental variable, one obtains a consistent estimator of the APE \citep[see, among others,][]{wooldridge2003,florens2008,wooldridge2008,heckman2010,masten2016,masten2017}. 

In this work, we instead consider identification and estimation of the random coefficient model in \eqref{eq:mainmodel} where valid instrumental variables may not be available, and the researcher would like to obtain a consistent estimator of the APE. We restrict our attention to the case where $X$ is a continuous random vector.

For simplicity, we consider first the case in which all elements of $X_i$ are correlated with the vector of random coefficients, and we defer the discussion of a more general model which includes additional exogenous controls. The random error terms, $\eta_i$ and $\varepsilon_i$, are such that $E \left[\eta_i \vert X_i \right] = a(X_i)$, and $E \left[\varepsilon_i \vert X_i \right] = b(X_i)$, where $b(X_i) = \left[b_1(X_i), \dots, b_p(X_i) \right]^\prime$, a $p \times 1$ vector of functions. Notice that, by construction, these functions have mean equal to zero, as
\[
E \left[ \eta_i \right] = E \left[ E \left[ \eta_i \vert X_i \right] \right] = E \left[ a(X_i) \right]= 0, 
\]
using the law of iterated expectations. Similarly, $E \left[ b(X_i)\right]=0$. 

With this notation, we can rewrite the model in \eqref{eq:mainmodel} as,
\[
Y_i = \alpha + a(X_i) + X_i \left( \beta + b(X_i) \right) + U_i,
\]
where
\[
U_i = \left( \eta_i - a(X_i) \right) + X_i\left( \varepsilon_i - b(X_i)\right), \text{ and } E\left[ U_i \vert X_i\right] = 0.
\]

We show that, when the function $a(X_i)$ is allowed to vary with $X_i$, it is not possible to identify any relevant policy parameter of the model without resorting to instruments. However, if $\eta_i$ is not correlated with $X_i$, then it is possible to recover the APE and the vector of conditional means of the random slopes. When $p=1$, this result is achieved without any additional conditions on the model parameters. On the contrary, when $p>1$, we need to impose further functional restrictions on $b(X_i)$, i.e. additivity. 

For estimation, we use existing semi/nonparametric techniques that are appropriately tailored to match the identification assumptions. We propose a semi-parametric series estimator that is flexible enough to easily impose our restrictions \citep{newey1997,chen2007}. In particular, notice that the reduced form of our model is cast as a partially varying (or semi-varying) coefficient model \citep[see][among others]{hastie1993,fan1999,li2002,ahmad2005,fan2005}. Our analysis thus builds on this rich statistical literature. Differently from existing work, however, the functional coefficients depend on all the regressors in the model, and, therefore, our identification and estimation analysis is, to some extent, innovative \citep[see also][for a similar generalization of the varying coefficient model]{lee2012}.   

While the assumption that the random intercept is not correlated with the regressors is unlikely to hold in many economic models (like the model of return to schooling mentioned above), empirical researchers could fruitfully apply our identification and estimation strategy in short panel data and, similarly, in difference-in-difference estimation with a continuous treatment, when the treatment effect is heterogeneous across the population. As a matter of fact, when the time dimension is small, it usually cannot be used as an additional source of variation to help identify the APE. However, differencing can remove the intercept from the regression model, and transform it into a framework that fits our assumptions. 

Consider, for simplicity, the case in which $p = 1$, and we have two time periods, $t = 1,2$. We write
\begin{align*}
Y_{i 1}  =& \alpha_i + X_{i 1}\beta_i + e_{i1},\\
Y_{i 2}  =& \alpha_i + X_{i 2}\beta_i + e_{i2},
\end{align*}

where we have omitted a trend component, and $\beta_i = \beta + \varepsilon_i$. As in \citet{wooldridge2005}, we assume that 
\begin{equation*} \label{eq:strongexo}
E \left[ e_{it}\vert X_{i 1},X_{i 2},\alpha_i,\beta_i\right] = 0, \text{ with } t =1,2,
\end{equation*}
which is a standard strong exogeneity condition in unobserved-effects models. Although this assumption restricts the possibility of lagged dependent variables, it leaves the correlation between $\lbrace X_{i1},X_{i2}\rbrace$, and $\lbrace \alpha_i,\beta_i \rbrace$ unspecified. 

We then take first differences to eliminate the random intercept, and obtain
\[
\Delta Y_i =  \Delta  X_{i}\beta_i + \Delta  e_{i},
\]
with $\Delta Y_i = Y_{i 2} - Y_{i 1}$, $\Delta X_i = X_{i 2} - X_{i 1}$, and $\Delta  e_{i} = e_{i 2} - e_{i 1}$. Because of the strict exogeneity assumption, we have that $E\left[ \Delta  e_{i} \vert \Delta  X_{i} \right] = 0$. Moreover, in this model, the random intercept is constant and one could use our approach to identify the APE and the conditional expectation, $E \left[ \varepsilon_i \vert  \Delta  X_{i} \right]$, without resorting to external sources of exogenous variation (i.e. instrumental variables). 

\citet{bleakley2010} exploits organized effort to eradicate malaria as a natural experiment to assess the effect of malaria on economic outcomes in the US and South America (Brazil, Mexico and Colombia). He uses a difference-in-difference approach in which he compares trends in adult income by birth cohort in areas that may or may not have experienced reductions in the incidence of malaria because of the eradication campaigns. In his framework, treatment is a continuous variable. Regions with high levels of pre-eradication malaria would potentially benefit more from the campaign. While regions that are Malaria-free would not benefit from the program and are used as a control group. In a model with constant treatment effect, he finds that, relative to non-malarious areas, cohorts born after eradication had higher income as adults than the preceding generation.

We replicate his analysis with Colombian data by allowing the treatment effect to be heterogeneous across regions. We first argue that, if the treatment effect is heterogeneous and negatively correlated with pre-eradication levels of malaria, then a simple OLS estimator may be downward biased. Using our approach, we confirm that this negative correlation is plausible and we find a larger effect of eradication campaigns on economic outcomes.  

Our contention is that our empirical strategy is applicable to any linear model with heterogeneous responses and continuous treatment, whenever the treatment and the outcome are observed at two time periods, and the strong exogeneity condition holds.

The paper is structured as follows. We discuss identification and estimation of a semiparametric estimator of this model in Sections \ref{sec:iden} and \ref{sec:est}, respectively. Section \ref{sec:montecarlo} presents the results of a large simulation study, in which we show that our estimator behaves well under several data generating processes. The empirical example of Section \ref{sec:empiricalapp} concludes. 

\subsection*{Notations} In the following, we let $\mathbbm{L}^2$, the space of square integrable functions with respect to the Lebesgue measure. For a real valued function $b_j \in \mathbbm{L}^2$, we let $\Vert b_j \Vert_2$ be the $\mathbb{L}^2$-norm; and, for a vector valued function $b = [b_1,\dots,b_p]^\prime$, we let $\Vert b \Vert_2 = \sqrt{\sum_{j=1}^p \Vert b_j \Vert^2_2}$. Similarly, we let $\sup_x \vert b_j(x) \vert$, to be the supremum norm of the function $b_j$; and for the vector valued function $b$, we let $\sup_x \vert b(x) \vert = \max_j \sup_x \vert b_j(x) \vert$. We also let $\Vert \cdot \Vert_{\ell^2}$ to be the Euclidean norm for vectors, and the induced operator norm for matrices. Finally, for a triplet of random variables $X_1$, $X_2$ and $X_3$, we denote $X_1 \upmodels X_2 \Vert X_3$, if $X_1$ and $X_2$ are independent given $X_3$.

\section{Identification} \label{sec:iden}

In the following, we let $X_{-l}$ be the vector $X$ where its $l^{th}$ component has been removed.

Notice that writing $\tilde{a} (x) = a(x) - x_j \gamma$, for some $\gamma \in \R$, and $j = 1,\dots,p$, would allow one to redefine $\tilde{\beta}_j = \beta_j + \gamma$, so that $a(x)$ and $\beta$ are not separately identified without additional restrictions. Similarly, for $\theta \in \R$, one could write $\tilde{b}_l (x) = b_l(x) + x_j \theta$, and $\tilde{b}_j (x) = b_j(x) - x_l \theta$, for every $j,l = 1, \dots p$, with $j\neq l$. This implies that the parts of $b_l(x)$ which are linear in $X_{-l}$, for all $l = 1,\dots,p$, are not separately identified. Finally, suppose that for some $j \in 1,\dots,p$, $b_j(x)$ is a purely multiplicative function. Without loss of generality, it can be written as
\[
b_j(x) = \prod_{l = 1}^p x^{d_l}_{l}, 
\]
with $d_l \in \mathbb{Z}$, for $l = 1,\dots,p$. Then, for some $l^\prime \in 1,\dots,p$, $l^\prime \neq j$, one could redefine
\[
\tilde{b}_{l^\prime}(x) = b_{l^\prime}(x) - x^{d_{l^\prime}-1}_{l^\prime} x^{d_j + 1}_{j} \prod_{l = 1, l \neq l^\prime,j}^p x^{d_l}_{l},
\]
in a way that $b_j$ and $b_{l^\prime}$ are not separately identified.

Thus, to obtain point identification of the parameters of the model, we impose the following additional restrictions. For simplicity, we adopt the notation $W_i$, to indicate the vector of covariates that includes a constant term and all the $p(p-1)/2$ linear interaction terms between the components of the random vector $X_i$. That is,
\[
W_i = \begin{bmatrix} 1 & X_i & \lbrace X_{ji} X_{li}, j,l = 1,\dots,p, l>j \rbrace \end{bmatrix}
\]
a row vector of dimension $p(p + 1)/2 + 1$. We further let $X_0$ to be the constant component of the model. 

\begin{assumption}~ \label{ass:identification1}
\begin{itemize}
\item[(i)] The functions $\lbrace a(x),b(x) \rbrace$ are infinitely differentiable at $0$. 
\item[(ii)] We let $a(x) = a_{00} + \sum_{j = 0}^{p} \sum_{l = 1, l > j}^{p} X_j X_l a_{jl}$, where $a_{00}$ and $\lbrace a_{jl}, j,l = 0,\dots,p,l > j\rbrace$ are constant parameters.
\item[(iii)] The matrix $E \left[ W^\prime_i W_i \right]$ has full rank.
\item[(iv)] The smallest eigenvalue of the matrix $E \left[ X^\prime_i X_i \vert X_{j i} = x_j\right]$ is bounded away from $0$, for all $j = 1,\dots,p$. 
\item[(v)] $E \left[ X_{li} b_{j}(X_i)\right] = 0$, for all $j,l = 1,\dots,p$ and $l \neq j$.
\item[(vi)] $b_j(X_i) = \sum_{l=1}^p b_{j,l}(X_{li})$, and for $\lbrace b_{j,l},  l = 1,\dots,p \rbrace$ not trivial, we cannot have $b_j(x) = 0$ for all $x$ in the support of $X$, for $j = 1,\dots,p$. 
\end{itemize}
\end{assumption}

Part (i) of this proposition imposes sufficient regularity on the vector of conditional means. Given the structure of the model, this requirement allows us to write the Maclaurin series of the unknown functions. We let $\mathcal{M}$ to be the class of functions which satisfy Assumption \ref{ass:identification1}(i). The functional coefficients $\lbrace a(x), b(x)\rbrace$ capture the heterogeneity in the marginal effect of $X$, so that the role of parts (ii), and (v) is to assign the \textit{linear} and the first-order \textit{interaction} effects to the functional intercept (part ii); while all other nonlinear effects are captured by the functional slopes (part v). Parts (iii) and (iv) are regularity conditions on the matrix of second moments. Recall that the vector of regressors $X_i$ does not contain a constant term, so that part (iv) is usually satisfied \citep[see][for a similar assumption]{lee2012}. Part (vi) restricts the functions $b_{j}$ to be additive, and imposes a standard no-concurtivity condition on the additive components. As noted above, this condition is essential to identify the interaction effects between each covariate and the nonlinear functional coefficients. Although it implies additional restrictions on the correlation structure between the random coefficients and the regressors, it is milder than some of the parametric assumptions used in this literature \citep{wooldridge2005}. Furthermore, empirical models in the social sciences often focus on the case of a scalar endogenous covariate, and the assumption of additivity becomes irrelevant in the latter case (see Example \ref{rem:scalarmod} below).

\begin{remark}\label{remark:identification}
In an attempt to clarify the implications of Assumption \ref{ass:identification1}(i), consider the simple case in which $X \in \mathbb{R}$. Assume $E(1/X^\kappa)$, for $\kappa > 0$ exists, and let 
\[
b(x)= b\left[\frac{1}{x^\kappa} - E\left( \frac{1}{x^\kappa} \right) \right],
\]
with $b$ a constant. Notice that this function fails to satisfy our restrictions, as it is not differentiable at $0$ (as a matter of fact, it is even not defined at $0$). 
\end{remark}

Under Assumption \ref{ass:identification1}, we can rewrite our model as follows
\begin{equation} \label{eq:multmodel}
Y_i =  \alpha + a_{00} + \sum_{j = 1}^p  X_{ji} \left( a_{0j} + \beta_j \right) + \sum_{j = 1}^p \sum_{l=1,l > j}^p  X_{ji} X_{li} a_{jl} + X_i b(X_i) + U_i.
\end{equation}

To simplify notations, we let $\delta = \lbrace \alpha + a_{00}, a_{0j} + \beta_j, a_{jl}, j = 1,\dots,p, l > j\rbrace^\prime$ to denote the vector of coefficients associated with the parametric components of the model. Similarly, we use a different representation of the nonparametric part of our model by collecting those functional coefficients that depend on the same continuous covariate. This representation is useful to derive the identification results and provides a simple estimation strategy. We thus let
\[
X_i b(X_i) = X_i \sum_{j = 1}^p b_{j}^\ast (X_{j i}),
\]
where $b_{j}^\ast (X_{j i}) = \left[ b_{1,j} (X_{j i}), \dots, b_{p,j} (X_{j i})\right]^\prime$ is a $p\times 1$ column vector which only includes the components of $b(X_i)$ that depend on $X_{j i}$, for each $j = 1,\dots,p$; and the summation applies rowwise. We thus write

\begin{equation} \label{eq:estmodel}
Y_i =  W_i \delta + X_i b^\ast (X_{i})+U_{i},
\end{equation}

where $b^\ast \equiv b$ is a $p \times 1$ vector of functions.

We provide two results. The first result is a \textit{negative} result: it excludes the possibility of identifying any relevant policy parameter in our setting. The second result is instead a \textit{positive} result. When we assume that the random intercept is not correlated with the regressors, we can identify the APE of the model under our assumptions. 

\begin{proposition}~ \label{prop:identification}
Under Assumption \ref{ass:identification1}, the vector of parameters, $\delta$, and the vector of additive functional coefficients, $b^\ast \in \mathcal{M}$, are identified. However, the vector of Average Partial Effects, $\beta$, is not identified. 
\end{proposition}

A proof of this Proposition is given in Appendix. The first statement of the theorem suggests that even by taking a finite order approximation for the function $a(X_i)$, its parameters are confounded with the Average Partial Effects. This makes identification of the latter impossible. In the simplest possible case when $p = 1$, with $a(X_i) = a_0 + X_i a_1$, and $b = 0$, almost surely, one could only hope to identify the parameters $\lbrace a_0, \beta + a_1 \rbrace$. This case is tantamount to the simple omitted variable example that can be found in many undergraduate statistics and econometrics textbooks.

We now add to Assumption \ref{ass:identification1}(ii) the following additional condition.
\begin{assumpB} \label{ass:identification21b}
The parameters $\lbrace a_{00},a_{0j}, j = 1,\dots,p\rbrace$ are identically equal to $0$.
\end{assumpB}

While this Assumption may appear ad hoc, it is tantamount to assume that $E\left[ \eta_i \vert X_i\right] = 0$, and to assign the linear components of $b^\ast$, which are not separately identified by Assumption \ref{ass:identification1}(v), to the function $a$. This implies the following. 

\begin{proposition}\label{prop:identification1}
Let Assumptions \ref{ass:identification1} and \ref{ass:identification21b} hold. We obtain that $\delta = \lbrace \alpha, \beta_j, a_{jl}, j,l = 1,\dots,p, l > j \rbrace^\prime$ in Proposition \ref{prop:identification}.
\end{proposition}

This proposition is given without proof, as it easily follows from that of Proposition \ref{prop:identification} above. It shows that, when we rule out the correlation between the intercept and the regressors, we are able to recover the Average Partial Effects without resorting to other sources of exogenous variation. The vector of conditional means $b^\ast \in \mathcal{M}$ is, in general, not uniquely identified. In the scalar case, however, the condition in Assumption \ref{ass:identification1}(v) becomes irrelevant and we are able to uniquely identify both the APE and the conditional mean $b^\ast$, as the following example shows. 

\begin{example} \label{rem:scalarmod}
When $p = 1$, and $a = 0$ almost surely, then it is possible to identify both the Average Partial Effect $\beta$, and the conditional mean function, $b(X_i) \equiv b^\ast(X_i)$, without imposing additional conditions \citep{heckman1998}.\footnote{We thank an anonymous referee for suggesting this example.} As $E\left[ b(X_i) \right] = 0$ by construction, then 
\[
Y_i = \alpha + X_i \beta + X_i b(X_i) + U_i.
\]
The parameters of this model are identified as long as
\begin{equation} \label{eq:singlevarmod}
\alpha + X_i \beta + X_i b(X_i) = 0 \Rightarrow \alpha =  \beta = b = 0, 
\end{equation}
where all equalities are intended almost surely. Notice that the function $b$ must satisfy Assumption \ref{ass:identification1}(i) and must be centered. Thus 
\begin{equation}\label{eq:identifid}
X_i b(X_i) = -\alpha - X_i \beta.
\end{equation}
Thus, the function $b(X_i)$ that satisfies this restriction must be a centered linear function. Notice that Assumption \ref{ass:identification1}(i) excludes the possibility that we might have
\[
b(x) = -\frac{b}{x^\kappa},
\]
for constants $b$ and $\kappa > 0$, as explained in Remark \ref{remark:identification}. Thus, the identity in \eqref{eq:identifid} directly gives $\alpha = 0$, because the function on the left-hand-side does not have an intercept and $X_i$ is almost surely not a constant (Assumption \ref{ass:identification1}(iii)). Therefore, the function $b(X_i)$ must satisfy
\[
b(X_i) = -\beta, 
\]
a constant. However, the expectation of $b(x)$ is equal to $0$ by construction and the only constant function with mean $0$ is the trivial function. The implication in \eqref{eq:singlevarmod} follows.
\end{example}

\begin{remark}[Short panel]
This identification approach can be adopted in short panel data models with correlated random coefficients, as detailed in the introduction, provided the strong exogeneity condition in \eqref{eq:strongexo} is satisfied.
\end{remark}

\begin{example}[Bivariate case]
We show in the bivariate case ($X\in \R^2$) how the structural parameters of the model are related to the reduced form parameters of the semi-varying coefficient model. Consider the following model, 
\[
Y_i = \alpha_{i} + X_{1i} \beta_{1i}  +  X_{2i} \beta_{2i},
\]
where we take 
\begin{align*}
E \left[ \alpha_{i} \vert  X_i \right] =& \alpha + E \left[ \eta_{i} \vert X_i \right] = 0\\
E \left[ \beta_{1i}\vert  X_i \right]   =& \beta_1 + E \left[ \varepsilon_{1i} \vert X_i \right] = X_{1i} + X_{2i}\\ 
E \left[ \beta_{2i} \vert  X_i \right] =& \beta_2 + E \left[ \varepsilon_{2i} \vert X_i \right] = X_{1i} - X_{2i}.
\end{align*}
For simplicity, we assume that $X_1$ and $X_2$ and stochastically independent, with $E \left[  X_{1i}\right] = E \left[ X_{2i}\right] = 0$. Using the reduced form model, we have that,
\begin{align*}
E \left[ Y_i \vert X_i \right] =& X_{1i} \left( X_{1i} + X_{2i} \right) +  X_{2i} \left( X_{1i} - X_{2i} \right)  \\
=& 2 X_{1i} X_{2i} + X^2_{1i} - X^2_{2i},
\end{align*}
so that $\delta_0 = 0$, $\delta_1 = \delta_2 = 0$, $\delta_3 = 2$, $b^\ast_{1,1} \left( X_{1i} \right) = X_{1i}$; $b^\ast_{2,1} \left( X_{1i} \right) = 0$; $b^\ast_{1,2} \left( X_{2i} \right) = 0$; and $b^\ast_{2,2} \left( X_{2i} \right) = -X_{2i}$. 

Notice that $E \left[ b^\ast_1 \left( X_i \right)  \right] = E \left[ b^\ast_2 \left( X_i \right)  \right] = 0$. Similarly, 
\begin{align*}
E & \left[ X_{2i} b^\ast_{1,1} \left( X_{1i} \right)  \right] = E \left[ X_{2i} X_{1i}  \right] = 0\\
E & \left[ X_{1i} b^\ast_{2,2}\left( X_{2i} \right)  \right] = -E \left[ X_{1i} X_{2i}  \right] = 0,
\end{align*}
because of the independence and the zero mean assumptions.
\end{example}

\section{Estimation} \label{sec:est}

We work with a sample $\lbrace (Y_i,X_i), i=1,\dots,n\rbrace$ of independent and identically distributed (i.i.d.) observations from the joint distribution of the random vector $(Y,X)$. 

Notice that the model as written in equation \eqref{eq:estmodel} is a partially varying (or semi-varying) coefficient model \citep[see][among others]{hastie1993,fan1999,li2002,ahmad2005,fan2005}. Here, we focus on a flexible series estimator of this semi-parametric model, and we defer the study of a kernel based estimator to further research. 

Our specification slightly differs from the standard framework of varying and semi-varying coefficient models. In this class of models, authors usually distinguish variables that enter linearly (regressors), and those that affect only the functional coefficient (covariates). These usually have no elements in common. In our case, the functional coefficients depend on all the regressors in the model, and, therefore, regressors and covariates perfectly overlap \citep[see also][for a similar generalization of the varying coefficient model]{lee2012}. 
The vector of parameters, $\lbrace b^\ast,\delta \rbrace$, satisfies the following system of conditional moment restrictions
\begin{align*}
E & \left[ X^\prime_i \left( Y_i - W_i \delta - X_i \sum_{j = 1}^p b_{j}^\ast (X_{j i}) \right) \vert X_{j i} = x_j \right] = 0,\text{ for } j = 1,\dots,p\\
E & \left[ W^\prime_i \left( Y_i - W_i \delta - X_i \sum_{j = 1}^p b_{j}^\ast (X_{j i}) \right) \right] = 0,
\end{align*}
subject to $E \left[ b^\ast_{j}(X_{j i}) \right] = 0$, $E \left[ X_{-j^\prime i} b^\ast_{ j^\prime,j}(X_{j i}) \right] = 0$, for all $j,j^\prime = 1,\dots,p$.

\subsection{A Semiparametric Sieve Estimator} We take the following family of basis functions $\psi_{K}(\cdot) = \left\lbrace \psi_{1,K}(\cdot),\dots,\psi_{K,K}(\cdot)\right\rbrace$ of dimension $K >0$ (e.g., B-splines, polynomials). We let $\tilde{\psi}_{j^\prime j K}(\cdot)$ be the transformation of $\psi_{K}(\cdot)$ that embeds the restrictions of Assumption \ref{ass:identification1} on the $j^\prime$-th component of the functional vector $b^\ast_{j}$, with $j,j^\prime = 1,\dots,p$. That is, $E\left[ \tilde{\psi}_{j j K}(X_{ji}) \right]  = E\left[ \tilde{\psi}_{j^\prime j K}(X_{ji}) \right] = 0$, and $E\left[ X_{-j^\prime i}\tilde{\psi}_{j^\prime j K}(X_{ji}) \right] = 0$, for all $j,j^\prime,l = 1,\dots,p$.

Finally, abusing notation slightly, let
\[
\tilde{\Psi}_j^\oplus (\xi) = \left[ \begin{array}{ccccc}
\tilde{\psi}_{1j K}(\xi) & \multicolumn{4}{c}{0_{ 1 \times (p-1)K}} \\
0_{1 \times K} & \tilde{\psi}_{2 jK}(\xi) & \multicolumn{3}{c}{0_{ 1 \times (p-2)K}} \\
\multicolumn{2}{c}{0_{1 \times 2K}} & \tilde{\psi}_{3 j K}(\xi)  & \multicolumn{2}{c}{0_{ J \times (d-2)}} \\
\multicolumn{3}{c}{\ddots}  & \ddots &  \ddots \\
\multicolumn{4}{c}{0_{ 1 \times (p-1)K}} &  \tilde{\psi}_{p j K}(\xi)   \end{array}\right],
\]
the $p \times pK$ direct sum at a given point $\xi$ of the vectors $\tilde{\psi}_{j^\prime j K}(\xi)$ for $j^\prime = 1,\dots,p$. For ease of notations, we take the dimension of the basis function, $K$, to be the same across all components of $X$, but one could easily accommodate a different smoothing parameter for each $X$. 

We make the following assumption about the additive functional coefficients.

\begin{assumption}\label{ass:smoothness}
$\sum_{l = 1}^p E\left[ X^2_{li} b_l^2(X_i)\right] < \infty$, and $\lbrace b^\ast_{j^\prime,j}, j,j^\prime = 1,\dots,p \rbrace$ belong to some class of smoothness $\mathcal{B}$. 
\end{assumption}

Usually, $\mathcal{B}$ is taken to be some H\"older or Sobolev class of functions \citep[see][for additional examples]{chen2007}. However, we do not restrict it further. One could tailor this class to the specific empirical problem at hand, and to the approximation properties of the basis functions used. For instance, polynomials may or may not have optimal uniform properties depending on the specification of $\mathcal{B}$ \citep{newey1997,belloni2014}.

We presume that the linear span of $\psi_K(\cdot)$ is dense in $\mathcal{B}$. We can therefore approximate each vector of functions $b^\ast_j$ as follows
\[
b^{\ast,K}_j(\xi) = \tilde{\Psi}_j^\oplus (\xi) \pi_{j} , \quad \forall j = 1,\dots,p,
\]
with $\pi_j$ being a $pK$-vector of generalized Fourier coefficients.

Finally, we let 
\[
S(X_i) = \left[ X_i \tilde{\Psi}_1^\oplus (X_{1i}), \dots, X_i \tilde{\Psi}_p^\oplus (X_{pi}) \right].
\]

The $p^2 K$-column vector of coefficients $\pi = \lbrace \pi_{j}, j = 1,\dots,p \rbrace$ and the finite dimensional parameter $\delta$ thus satisfy the following system of unconditional moment restrictions
\[
E\left[ \left(W_i , S (X_i ) \right)^\prime  \left( Y_i - W_i \delta - S (X_i) \pi \right)\right] = 0,
\]
where the expectation is taken with respect to the joint distribution of $\left(Y,X\right)$.

Starting from this moment condition, we implement a profile least-square procedure and we study the properties of an estimator of the finite dimensional parameter $\delta$ and of the entire additive structure $b^\ast$, for $j =1,\dots,p$ \citep{stone1985,ahmad2005,fan2005}. 

To this end, let $\PM = E \left[ S (X_i )^\prime S (X_i ) \right]$, and 
\begin{align}
\pi =& \PM^{-1} \left( E \left[ S(X_i)^\prime Y_i\right]  - E \left[ S(X_i)^\prime W_i  \right] \delta  \right),\label{eq:pi0}\\
\delta =& \left( E \left[ W_i^\prime W_i\right] -  E \left[ W_i^\prime S (X_i ) \right] \PM^{-1}  E \left[  S (X_i )^\prime W_i \right] \right)^{-1} \times \notag\\
  & \left( E \left[ W_i^\prime Y_i\right]  - E \left[ W_i^\prime S (X_i ) \right] \PM^{-1}  E \left[  S (X_i )^\prime Y_i \right] \right)\label{eq:delta0}.
\end{align}

Our estimators of $\delta$ and $\pi$ are simply the sample counterparts of equations \eqref{eq:pi0} and \eqref{eq:delta0}. Letting, $\SM_n = \left[ S(X_1) \dots S(X_n)\right]$, and $\hat{\PM}_n = \frac{1}{n}\sum_{i = 1}^n S(X_i)^\prime S(X_i)$, we have
\begin{align}
\hat{\pi} =& \hat{\PM}_n^{-} \SM^\prime_n \left( \mathbf{Y}_n  - \mathbf{W}_n \hat{\delta}\right)/n,\label{eq:pihat}\\
\hat{\delta} =& \left(\frac{\mathbf{W}_n^\prime \mathbf{W}_n}{n}-  \frac{\mathbf{W}_n^\prime \SM_n}{n} \hat{\PM}_n^{-}  \frac{\SM_n^\prime \mathbf{W}_n}{n} \right)^{-1}\left( \frac{\mathbf{W}_n^\prime \mathbf{Y}_n}{n}  - \frac{\mathbf{W}_n^\prime \SM_n}{n} \hat{\PM}^{-} \frac{\SM_n^\prime \mathbf{Y}_n}{n} \right)\label{eq:deltahat},
\end{align}
where $(\cdot)^{-}$ denotes the generalized inverse of a matrix, $\mathbf{Y}_n$ and $\mathbf{W}_n$ are, respectively, a $n \times 1$ vector of sample observations of the response variable $Y$, and a $n \times p(p+1)/2 + 1$ matrix of sample observations of the independent variable $W$. 

For each $j = 1,\dots,p$, we thus have
\[
\hat{b}^{\ast,K}_j(\xi_j) = \tilde{\Psi}_j^\oplus (\xi)\hat{\pi}_{j}, 
\]
a $p \times 1$ vector of functions, and 
\[
\hat{b}^{\ast,K}(\xi)  = \sum_{j = 1}^p \hat{b}^{\ast,K}_j(\xi_j)  = \tilde{\mathbf{\Psi}}^\oplus(\xi) \hat{\pi},
\]
with 
\[
\tilde{\mathbf{\Psi}}^\oplus(\xi) = \left[ \tilde{\Psi}_1^\oplus (\xi_{1}), \dots,\tilde{\Psi}_p^\oplus (\xi_{p}) \right],
\]
a $p\times p^2K$ matrix, and $\hat{\pi} = [\hat{\pi}^\prime_{1}, \dots, \hat{\pi}^\prime_p]^\prime$, a $p^2K + 1$ vector of coefficients. 

\subsection{Asymptotic properties} The asymptotic properties of these estimators are obtained following the framework in \citet{belloni2014} and \citet{chenchristensen2015}. We make the following additional Assumptions. 

\begin{assumption}~ \label{assest1}
\begin{itemize}
\item[(i)] The random vector $X \in \mathcal{X}$, a compact subset of $\mathbb{R}^p$. Its density function, $f_{X}(\cdot)$, is bounded away from $0$ and $\infty$ on its support.
\item[(ii)] $\sigma^2(X_i)= Var(U_i\vert X_i)$ is uniformly bounded away from $0$ and $\infty$ on $\mathcal{X}$. 
\end{itemize}
\end{assumption}

These assumptions are quite standard in the literature and do not deserve further explanation. In the following, to simplify notations, and without loss of generality, we take $\mathcal{X} = [0,1]^p$. One could relax Assumption \ref{assest1}(i) and allow for the support of the regressors to be unbounded at the cost of more involved proofs \citep{chenchristensen2013}.

\begin{assumption}~ \label{assest2}
Let $G_{\psi,j^\prime j} = E \left[\tilde{\psi}_{j^\prime j K} (X_{ji})^\prime \tilde{\psi}_{j^\prime j K}(X_{ji}) \right]$. 
\begin{itemize}
\item[(i)] The smallest eigenvalue of $G_{\psi,j^\prime j}$ is bounded away from zero, for all $K > 0$.
\item[(ii)] For all $j,j^\prime = 1 ,\dots,p$, 
\[
\sup_{\xi \in [0,1]} \Vert  \tilde{\psi}_{j^\prime jK}(\xi) G^{-1/2}_{\psi,j^\prime j} \Vert_{\ell^2} \leq \zeta_K.
\]
\item[(iii)] The smoothing parameter $K$ satisfies, $\zeta^2_K \log K/n = o(1)$, and $\zeta_K\sqrt{K/n} =O(1)$.
\item[(iv)] For some $\pi_{j^\prime j} \in \R^{K}$ and $b^\ast_{j^\prime,j} \in \mathcal{B}$
\[
\Vert b_{j^\prime,j}^{\ast,K} - b^\ast_{j^\prime,j} \Vert_2 = O_P \left( s_K\right),
\]
with $s_K \rightarrow 0$ as $K \rightarrow \infty$, for all $j,j^\prime = 1,\dots,p$. 
\end{itemize}
\end{assumption}

Parts (i) and (ii) are standard in this literature \citep[see][]{newey1997}. The former bounds the supremum of the Euclidean norm of the (orthornormalized) basis function. The latter requires that the basis functions are not too collinear. In part (ii), we implicitly assume that one uses the same basis function to approximate all the components of the functional vector $b$. Part (iii) describes restrictions on the growth of the basis functions to achieve consistency. Finally, part (iv) is an assumption on the order of approximation of the unknown function: this approximation becomes more accurate as the dimension of the basis functions diverges to infinity. Depending on the class $\mathcal{B}$ and on the properties of the basis functions used, we can obtain more explicit bounds on this approximation error. For instance, if $\mathcal{B}$ is a H\"older class of smoothness $s$, and we implement a B-spline estimator of the additive components, then $s_K = K^{-s/2}$. We make this assumption about the order of approximation for each component of the additive structure, but, as a consequence of Assumption \ref{ass:smoothness}, the same bound holds for the entire additive structure, and for all $j =1,\dots,p$. 

To derive the asymptotic properties of this estimator, we finally denote by $\mathcal{V}$ the class of functions of $X$ that can be written in the varying coefficient form $X_i \sum_{j = 1}^p b^\ast_{j}(X_{ji})$, with the functional components satisfying the restrictions of Assumption \ref{ass:identification1}. In particular, $b^\ast_{j} \in \mathcal{M}$, for all $j = 1,\dots,p$. For a random variable $W_i$, we denote as $E_{\mathcal{V}}\left[ W_i \right]$, the projection of $W_i$ onto $\mathcal{V}$, under the $\mathbb{L}^2$ norm.

Let us denote 
\begin{align}\label{eq:omegadef}
\Omega =& E\left[ \sigma^2 (X_i) \left( W_i - E_\mathcal{V}\left[ W_i \right]\right)^\prime  \left( W_i - E_\mathcal{V}\left[ W_i \right]\right)\right],  \\
\Phi   =& E\left[ \left( W_i - E_\mathcal{V}\left[ W_i \right]\right)^\prime  \left( W_i - E_\mathcal{V}\left[ W_i \right]\right) \right]
\end{align}

We can state the following result.
\begin{theorem}\label{thconvasnorm}
If $\sqrt{n} s_K \rightarrow 0$, then
\[
\sqrt{n} \left( \hat{\delta} - \delta \right) \overset{d}{\longrightarrow} N\left(0,\Phi^{-1} \Omega \Phi^{-1} \right).
\]
\end{theorem}

Notice that the estimator of the parameter $\delta$ reaches the semi-parametric efficiency bound under an additional assumption of homoskedasticity of the residual term \citep{ahmad2005,fan2005}. Homoskedasticity cannot be satisfied in our case, because of the way our model is derived. However, an efficient re-weighted estimator could be constructed along the lines of \citet{shen2014}. We do not pursue this extension here.

To obtain uniform rate of convergence for the nonparametric part of our model, we need further regularity conditions, which are given below \citep[see the more general theoretical development in][]{belloni2014}. 
\begin{assumption}~ \label{assest3}
There exists a constant $m > 2$, such that
\begin{itemize}
\item[(i)] $E \left[ U^m_i \vert X_i = x\right] < \infty$. 
\item[(ii)] $\zeta_K^{2m/(m-2)} \log K/n = O(1)$.
\item[(iii)] For some $\pi_{j^\prime j} \in \R^{K}$ and $b^\ast_{j^\prime ,j} \in \mathcal{B}$
\[
\sup_{\xi \in [0,1] }\vert b_{j^\prime ,j}^{\ast,K}(\xi) - b^\ast_{j^\prime ,j}(\xi) \vert = O_P \left( N_K s_K\right),
\]
with some $s_K \rightarrow 0$ as $K \rightarrow \infty$, and $N_K$, which depends on $\mathcal{B}$, for all $j,j^\prime = 1,\dots,p$. 
\end{itemize}
\end{assumption}
Part (i) provides a condition on the tails of the regression errors, while part (ii) is a condition on the growth of the basis functions. Part (iii) provides a uniform bound on the approximation error. The value of $N_K$ depends on the so-called Lebesgue constant \citep{devore1993}, and it is therefore defined as the Lebesgue factor in \citet{belloni2014}. For instance, the behavior of $N_K$ for polynomial series, when restricted to the class of continuous function on $[0,1]$ is such that $N_K \asymp K$. However, as discussed in \citet{belloni2014}, one could construct \textit{tailored} classes of functions for which $N_K \asymp \log K$, or even $N_K \asymp 1$. Depending on the specification of $\mathcal{B}$, it is thus possible to construct well-behaved uniform approximations with polynomial series. 

We have the following result.
\begin{theorem}\label{thconvproof}
Let Assumptions \ref{ass:identification1}, \ref{ass:smoothness}-\ref{assest2} hold. Then,
\[
\Vert \hat{b}^{\ast,K} - b^\ast \Vert_2 = O_P\left( \sqrt{\frac{K}{n}} + s_K \right). 
\]
If Assumption \ref{assest3} also holds, we further have,
\[
\sup_{\xi \in [0,1] }\vert \hat{b}^{\ast,K}(\xi) - b^\ast(\xi)\vert = O_P\left( \zeta_K \sqrt{\frac{\log K}{n}} + N_K s_K \right). 
\]
\end{theorem}

Rates for the estimation of the functional vector $b^\ast$ are the same as the univariate rate, when Assumption \ref{ass:smoothness} holds \citep{stone1985}. Depending on the choice of the function class $\mathcal{B}$ and on the approximating basis function, these uniform rates may be optimal, and we refer the reader to \citet{belloni2014} for further details. 

\subsection{Models with control variables} \label{subsec:exo}

In many relevant empirical cases, the researcher would like to add some control variables, $Z \in \mathbb{R}^q$ into the model.

When these control regressors are available, we rewrite our model as follows
\[
Y_i = \alpha_i + X_i \beta_i + Z_i \gamma_i,
\]
with
\begin{align*}
\alpha_i =& \alpha + \eta_i\\
\beta_i =& \beta + \varepsilon_i,\\
\gamma_i =& \gamma + \zeta_i,
\end{align*}
with $\eta_i$ and $\varepsilon_i$ defined as above, and $\zeta_i$ being a q-dimensional random vector of errors, possibly correlated with $X_i$, such that $E\left[ \zeta_i\right] <\infty$.

We directly make the following Assumption.

\begin{assumption}~\label{ass:control1}
$Z \upmodels \lbrace \varepsilon, \zeta \rbrace \Vert X$, and $E\left[ \eta_i \vert X_i,Z_i \right] = 0$. 
\end{assumption}

For simplicity, we directly assume that the constant term is exogenous. Furthermore, our assumption implies that the conditional expectation of the error terms only depends on the \textit{endogenous} regressors $X_i$, while it does not depend on $Z_i$. 

We thus let $c(X_i) = E \left[ \zeta_i \vert X_i\right]$. We can then write
\[
Y_i = \alpha + X_i \left( \beta + b(X_i) \right)  + Z_i \left( \gamma + c(X_i) \right) + U_i,
\]
where the definition of $U_i$ should be apparent and $E \left[ U_i \vert X_i,Z_i \right] = 0$, by construction and Assumption \ref{ass:control1}.

For identification and estimation, we need the following additional assumptions.

\begin{assumption}~ \label{ass:identificationcontrol1}
\begin{itemize}
\item[(i)] The marginal distributions of the control regressors $Z$ are either absolutely continuous w.r.t.\ the Lebesgue measure or they are discrete measures with finite support.
\item[(ii)] The matrix $E \left[ \left( W_i,  Z_i \right)^\prime \left( W_i, Z_i \right) \right]$ has full rank.
\item[(iii)] The smallest eigenvalue of the matrix $E\left[ \left( X_i,  Z_i \right)^\prime \left( X_i, Z_i \right)\vert X_{ji} = x_j\right]$, is almost surely bounded away from zero on $[0,1]$, for all $j = 1,\dots,p$.
\end{itemize}
\end{assumption}

Part $(i)$ imposes regularity conditions on the control variables $Z_i$. Part $(ii)$ and $(iii)$ provide regularity conditions on the unconditional and conditional design matrices. We finally have the following result.

\begin{proposition}\label{prop:controlvars}
Let Assumptions \ref{ass:identification1}(i), \ref{ass:identification21b}, \ref{ass:identification1}(v)-(vi), \ref{ass:control1}, and \ref{ass:identificationcontrol1} hold. The vector of parameters $\delta$, as defined in Proposition \ref{prop:identification1}, $\gamma$ and the conditional expectation functions, $b(x)$ and $c(x)$, are identified.
\end{proposition}

The proof follows from the proof of Proposition \ref{prop:identification} with the additional requirement that the full matrix of regressors has full column rank.

For estimation purposes, the model augmented to include additional control variables can be cast as a partially varying coefficient model, and it can be estimated under virtually the same conditions as above \citep{ahmad2005,lee2012}. The asymptotic properties of the resulting semi-parametric series estimator can thus be obtained in parallel with the results of the previous section. 

\section{Simulations} \label{sec:montecarlo}

We perform an extensive simulation study to assess the finite sample performance of our estimator. We take $n = \lbrace 100,250,500 \rbrace$ observations respectively, to assess the effect of an increasing sample size on the parametric part of our estimator. For a given design of the independent regressor, we construct $1000$ samples of the dependent variable. We use polynomial series to approximate the unknown nonparametric component of the model. We consider three cases
\begin{itemize}
\setlength{\itemindent}{.5in}
\item[\textbf{Design 1}:] Random Coefficients are correlated with the regressors, and we do not have available instruments. 
\item[\textbf{Design 2}:] Random Coefficients are not correlated with the regressors. 
\item[\textbf{Design 3}:] Random Coefficients are correlated with the regressors, and we have an instrumental variable. 
\end{itemize}
In Design 2, we compare our approach with a simple OLS. This gives us a consistent benchmark for the semi-parametric estimator. In Design 3, we also compare our estimator with a control function estimator. For Designs 1 and 2, we consider both univariate and bivariate $X$. For Design 3, we solely focus on the case where $X$ is univariate.  

In all Monte-Carlo studies with univariate independent variables, we have that $b \equiv b^\ast$, and we use the two notations concurrently. In all designs, the smoothing parameter (i.e., the order of the polynomial) is chosen by cross-validation \citep{hansen2012}. 

\subsection{Design 1}

\subsubsection{Univariate independent variable}

We consider the simple model with only one regressor
\[
Y_i = \alpha_{i} + X_i \beta_{i}, 
\]
where 
\[
X_i  \sim \mathcal{TN}[-1,1],
\]
where $\mathcal{TN}$ denotes a truncated normal distribution between $[-1,1]$. We leave $\alpha_i$ not dependent of $X_i$, and $\beta_i = \beta + \varepsilon_i$. 

For the endogenous case, we simulate directly from the reduced form of the model, 
\[
Y_i = \delta_{0} + X_i \delta_1 + X_i b(X_i) + U_i,
\]
where $\delta_{0} = 0$, and $b(X_i) = 2X_i^2 - 2E\left[X_i^2\right]$ and $\delta_1 = 2 E \left[ X_i^2 \right] = 0.582$. Recall that $U_i = X_i (\varepsilon_i - b(X_i))$. We take
\[
\varepsilon_i \vert X_i = x \sim  \mathcal{N}\left( b(x), \exp(0.5x)\right),
\]
so that $Var(U_i\vert X_i = x) = x^2 Var(\varepsilon_i\vert X_i = x)$. 

Table \ref{tabrmase1} below reports the MASE of the series estimator of $b(X_i)$.

This table indicates that our estimator performs well even for moderate sample sizes. 

\begin{table}[ht]
\centering
\begin{tabular}{ccc}
  \hline
100 & 250 & 500 \\ 
  \hline
0.15353 & 0.05610 & 0.02746 \\ 
   \hline
\end{tabular}
\caption{MASE of the nonparametric estimator} 
\label{tabrmase1}
\end{table}

\begin{figure}[!h]
\centering
\includegraphics[scale=0.6]{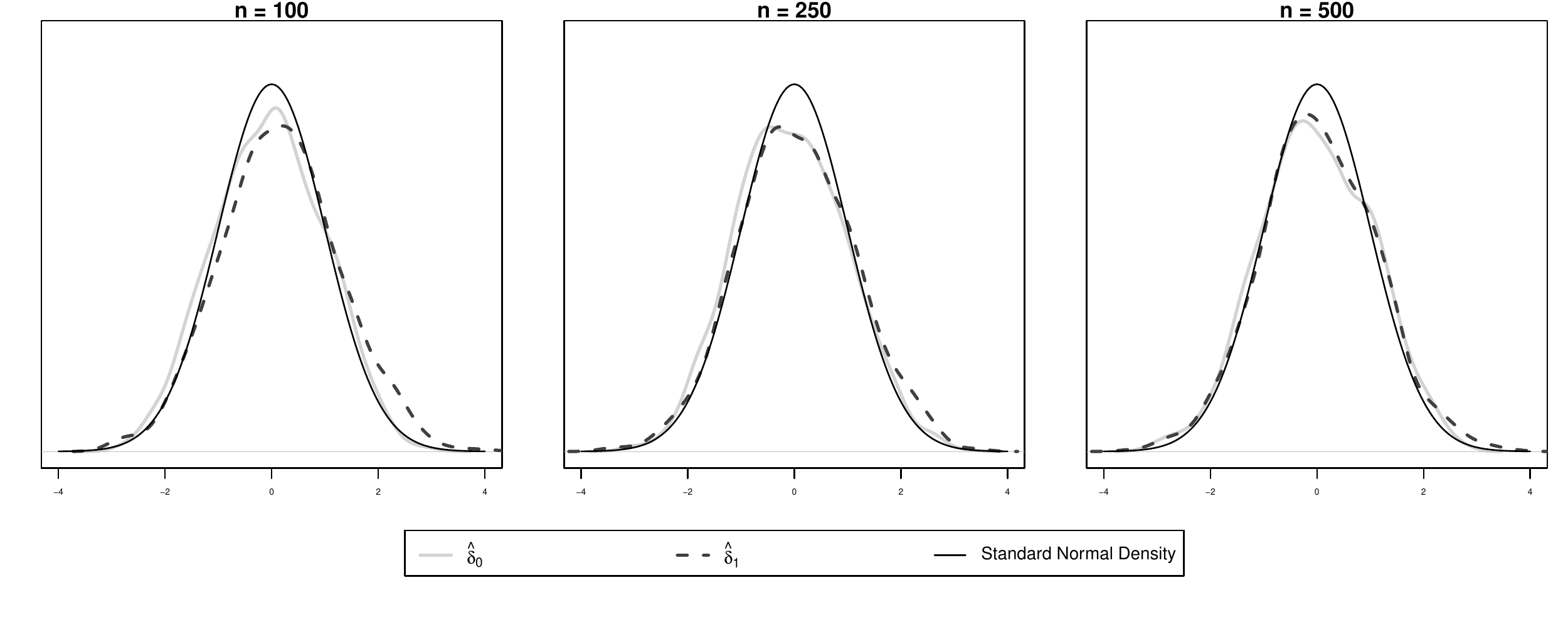}
\caption{Empirical densities of parametric estimators for increasing sample sizes}
\label{fig:fct_end1}
\end{figure}

We also report the empirical densities of the estimators of $\delta_0$ and $\delta_1$, centered, and normalized by the estimators of their asymptotic standard errors. In Figure \ref{fig:fct_end1}, the black solid line is the pdf of a standard normal variable. The light gray solid line is the empirical density of $\hat{\delta}_{0}$; and the dark grey dashed line is the empirical of $\hat{\delta}_{1}$. The sampling distributions of the parametric components of our model are converging to the standard normal distribution, as $n$ increases, corroborating the result of Theorem \ref{thconvasnorm}.

\subsubsection{Bivariate independent variable}

We consider the simple model with two regressors
\[
Y_i = \alpha_{i} + X_{1i} \beta_{1i} + X_{2i} \beta_{2i}, 
\]
with
\[
X_{i}  \sim \mathcal{TN}_2[-1,1],
\]
where $\mathcal{TN}_2$ is a truncated bivariate normal distribution between $[-1,1]$ \citep{botev2017}. We impose 
\begin{align*}
E \left[\alpha_{i} \vert X_i \right] =& \alpha + E \left[ \eta_{i} \vert X_i \right] = 0\\
E \left[\beta_{1i} \vert X_i \right] =& \beta_1 + E \left[ \varepsilon_{1i} \vert X_i \right] = 1.5 \left( X_{1i}^2 + X_{2i}^2 \right) \\ 
E \left[\beta_{2i} \vert X_i \right] =& \beta_2 + E \left[ \varepsilon_{2i} \vert X_i \right] = \exp \left(X_{1i}\right)  + \exp \left(X_{2i} \right).
\end{align*}
We simulate directly the reduced form of the model,
\begin{equation}\label{eq:estmodel2}
Y_i = \delta_{0} + X_{1i} \delta_1 + X_{2i} \delta_2 + X_{1i} X_{2i} \underbrace{\left( a_{12} + a_{21}\right)}_{\delta_3} + X_{1i} b_1(X) +  X_{2i} b_2(X) + U_i .
\end{equation}

Recall that $\delta_1$ and $a_{12}$, $\delta_2$ and $a_{21}$ can be obtained as the projection of the functions $E \left[ \varepsilon_{1i} \vert X_i \right]$ and $E \left[ \varepsilon_{2i} \vert X_i \right]$ onto the linear space spanned by $X_2$ and $X_1$, respectively. We then have, 
\begin{align*}
\delta_1 =& E\left[ 1.5 \left( X_{1i}^2 + X_{2i}^2 \right) \right] -\frac{Cov \left( X_{2i},1.5 \left( X_{1i}^2 + X_{2i}^2 \right) \right)}{Var \left(X_{2i} \right)}  E\left[ X_{2i }\right]  = 0.835\\
\delta_2 =& E\left[ \exp \left(X_{1i}\right)  + \exp \left(X_{2i} \right) \right] - \frac{Cov \left( X_{1i} , \exp \left(X_{1i}\right)  + \exp \left(X_{2i} \right) \right)}{Var \left(X_{1i} \right)} E\left[ X_{1i }\right] = 0.835\\
a_{12} =& \frac{Cov \left( X_{2i}, 1.5 \left( X_{1i}^2 + X_{2i}^2 \right) \right)}{Var \left(X_{2i} \right)} = 0\\
a_{21} =& \frac{Cov \left( X_{1i} ,  \exp \left(X_{1i}\right)  + \exp \left(X_{2i} \right) \right)}{Var \left(X_{1i} \right)} = 1.366.
\end{align*}

Hence, we have that $b_1(X_i) = 1.5 \left( X_{1i}^2 + X_{2i}^2 \right) - \delta_1$; and $b_2(X_i) =\exp \left(X_{1i}\right)  + \exp \left(X_{2i} \right) - \delta_2 - X_{1i} a_{21}$.

Finally, we take 
\[
\varepsilon_{i} \vert X_i = x \sim \mathcal{N} \begin{pmatrix} \begin{bmatrix} b_1(x) \\ b_2(x)\end{bmatrix} , \begin{bmatrix} exp(0.25(x_1 + x_2)) & \rho_{x} \\ \rho_{x} & 0.25(x_1 + x_2)^2 \end{bmatrix} \end{pmatrix},
\]
with $\rho_x = (x_1 x_2)\sqrt{0.25(x_1 + x_2)^2 \exp(0.25(x_1 + x_2))}$, in a way that 
\[
Var(U_i\vert X_i = x)  = x Var(\varepsilon_i \vert X_i = x) x^\prime.
\]

\begin{figure}[!h]
\includegraphics[scale=0.6]{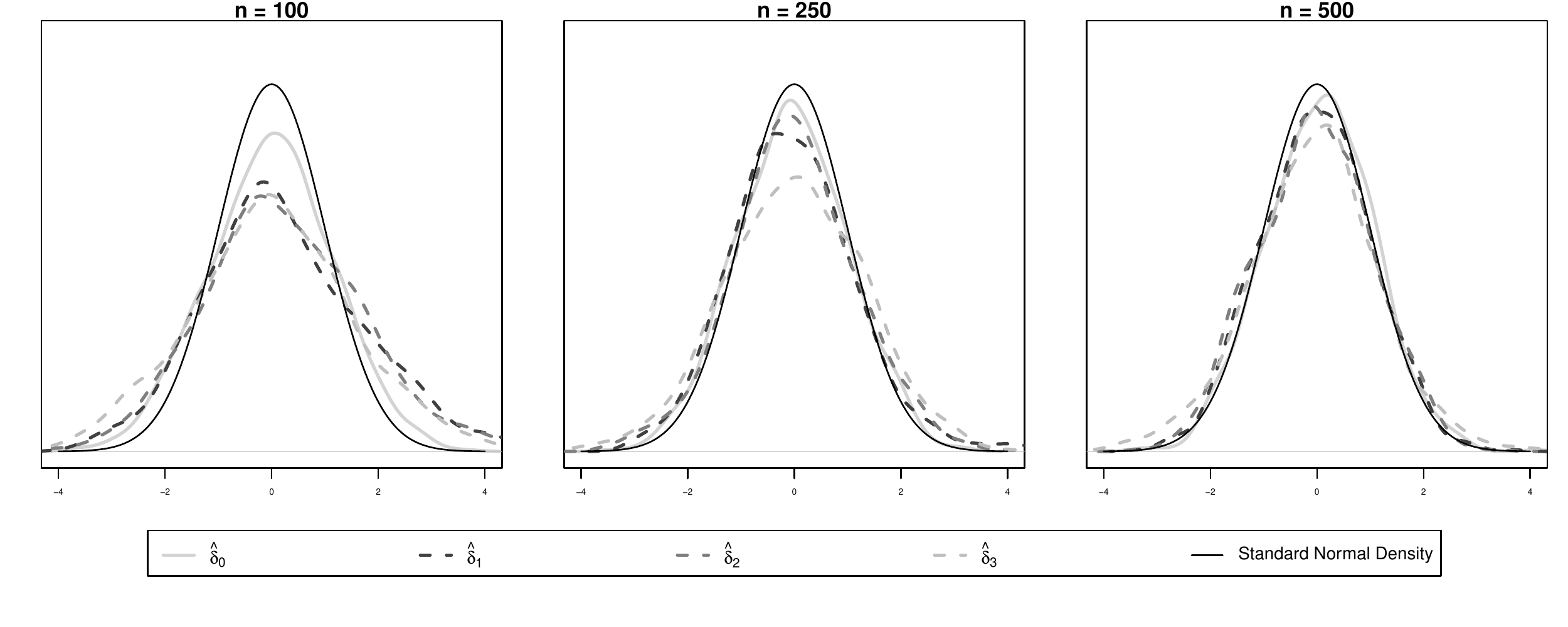}
\caption{Histogram of the density of the constant parameters}
\label{fig:fct_end2}
\end{figure}

We report below the Mean Average Squared Error (MASE) for the estimators of $b_1$ and $b_2$ (see Table \ref{tab11}) and the density as above for the estimators of the constant parameters of the model (see Figure \ref{fig:fct_end2}).

\begin{table}[ht]
\centering
\begin{tabular}{lccc}
  \hline
 & 100 & 250 & 500 \\ 
  \hline
$MASE(\hat{b}_1)$ & 0.315 & 0.113 & 0.054 \\ 
  $MASE(\hat{b}_2)$ & 0.272 & 0.111 & 0.053 \\ 
   \hline
\end{tabular}
\caption{MASE of $\hat{b}_1$ and $\hat{b}_2$ for increasing sample sizes.} 
\label{tab11}
\end{table}

The results in Table \ref{tab11} reveal that the estimators of both functions have good performance, and they both behave better with increasing sample size. In Figure \ref{fig:fct_end2}, the light grey solid line for estimated intercept and the dark grey, grey and light gray dashed lines for estimated $\delta_1$, $\delta_2$ and $\delta_3$ respectively, are all getting closer to the standard normal distribution (the continuous black line), in accordance with our asymptotic theory.

\subsection{Design 2}

\subsubsection{Univariate independent variable}

We simulate $X_{i}$ from a truncated normal distribution, as above. We take $\alpha = 0$, constant across all observations, and $\beta_{i} \sim \mathcal{N}(0.835,(0.835)^2)$. In this instance, we simulate directly from the structural model. 

We report below the RMSE of the OLS estimator and the RMSE of our semi-parametric estimator for $\alpha$ and $\beta$, where the latter parameter is an estimator of the mean of the distribution of $\beta_{i}$. 

\begin{table}[ht]
\centering
\begin{tabular}{l |ccc | ccc |}
  \hline
 & ~ & SNP & ~ & ~ & OLS & ~ \\ 
  \hline
Sample Size & 100 & 250 & 500 & 100 & 250 & 500 \\ 
   \hline
$\hat{\alpha}$ & 0.0171 & 0.0112 & 0.00793 & 0.0186 & 0.0118 & 0.0080 \\ 
  $\hat{\beta}$ & 0.0474 & 0.0290 & 0.02087 & 0.0479 & 0.0289 & 0.0207 \\ 
   \hline
\end{tabular}
\caption{RMSE of OLS versus semiparametric.} 
\label{tabols1}
\end{table}

Results are reported in Table \ref{tabols1}. Our estimator behaves very well compared to the simple OLS estimator. Also, the MASE for the nonparametric part of our model is equal to $0.0107$, $0.0036$, and $0.0014$, respectively. These results suggest that our estimator remains competitive even when the random coefficients are uncorrelated with the independent regressors.

\subsubsection{Bivariate independent variable}

The joint distribution of $X$ is a bivariate truncated normal in $[-1,1]^2$. We take $\alpha = 0$, constant across all observations, and $\beta_{1i} \sim \mathcal{N}(0.835,(0.835)^2)$, $\beta_{2i} \sim \mathcal{N}(2.291,(2.291)^2)$. We again simulate directly from the structural model. 

We report below the RMSE of the OLS estimator and the RMSE of our semi-parametric estimator for the triplet $\hat{\beta}_0$, $\hat{\beta}_1$, and $\hat{\beta}_2$, where the last two parameters estimate the mean of the distribution of $\beta_{1i}$ and $\beta_{2i}$, respectively. 

\begin{table}[ht]
\centering
\begin{tabular}{l |ccc | ccc |}
  \hline
 & ~ & SNP & ~ & ~ & OLS & ~ \\ 
  \hline
Sample Size & 100 & 250 & 500 & 100 & 250 & 500 \\ 
   \hline
$\hat{\alpha}$ & 0.123 & 0.0784 & 0.0579 & 0.0781 & 0.0472 & 0.0361 \\ 
  $\hat{\beta}_1$ & 0.226 & 0.1403 & 0.0973 & 0.1659 & 0.1046 & 0.0746 \\ 
  $\hat{\beta}_2$ & 0.215 & 0.1340 & 0.0916 & 0.2030 & 0.1262 & 0.0862 \\ 
   \hline
\end{tabular}
\caption{RMSE of OLS versus semiparametric.} 
\label{tabols2}
\end{table}

Results are reported in Table \ref{tabols2}. Our estimator compares well with OLS, although less favorably than in the univariate case. Our model is fully additive, and there is no curse of dimensionality. However, as it appears from the asymptotic properties, the variance of our estimator is larger in finite samples than the one of the standard OLS estimator, which entails a loss of efficiency. Finally, the MASE for the nonparametric part of our model is equal to $\lbrace 0.2198, 0.2321 \rbrace$, $\lbrace 0.0972, 0.087 \rbrace$, and $\lbrace 0.0427, 0.0446 \rbrace$, respectively. 

\subsection{Design 3}

For the last design, we generate an instrumental variable $Z \sim \mathcal{TN}[-1,1]$, and
\[
\begin{bmatrix} \eta_i \\ \beta_i \\  \zeta_i  \end{bmatrix} \sim \mathcal{N}_3 \left( \begin{bmatrix} 0 \\ 1 \\ 1 \end{bmatrix}, \begin{bmatrix} 1 &  0 & 0 \\ 0 & 1 & 0.4 \\ 0 & 0.4 & 1 \end{bmatrix} \right),
\]
with
\begin{align*}
X_i =& 1.5 Z_i + \zeta_i\\
Y_i =& \alpha + X_i \beta_i + 0.25\eta_i, 
\end{align*}
where $\eta_i$ is simulated independently of $\beta_i$ and $\zeta_i$. We take $\alpha = \delta_0 = 0$, and from above we have $\beta = \delta_1 = E \left[ \beta_{1i} \right] = 1$, in a way that we can take $\beta_i =\beta + \varepsilon_i$, where $\varepsilon_i$ is a zero mean error term. In this setting, we compare our estimator with both the OLS estimator (which is not consistent because of the correlation between $X_i$ and $\beta_i$), and a control function (CF) estimator \citep{wooldridge1997,heckman1998}. As control function, we employ the residuals from a first step linear regression of $X$ on $Z$. This control function is then included in a second step regression, along with an interaction term between the control function itself and $X$. 

We report below the RMSE of our semi-parametric estimator and the control function estimator for $\alpha$ and $\beta_1$. 

\begin{table}[ht]
\centering
\begin{tabular}{ll |ccc | ccc |ccc |}
  \hline
~ & ~ & ~ & SNP & ~ & ~ & OLS & ~ & ~ & CF & ~ \\ 
  \hline \multicolumn{2}{c |}{Sample Size} & 100 & 250 & 500 & 100 & 250 & 500 & 100 & 250 & 500 \\ \hline
~ & RMSE & 0.144 & 0.102 & 0.110 & 0.234 & 0.195 & 0.179 & 0.222 & 0.142 & 0.100 \\ 
  $\hat{\alpha}$ & BIAS & 0.0241 & 0.0149 & 0.0472 & 0.1520 & 0.1586 & 0.1614 & 0.0004 & 0.0066 & 0.0050 \\ 
  ~ & SE & 0.1422 & 0.1006 & 0.0996 & 0.1778 & 0.1143 & 0.0783 & 0.2224 & 0.1418 & 0.1002 \\ 
   \hline
~ & RMSE & 0.1592 & 0.0984 & 0.0776 & 0.3223 & 0.2759 & 0.2511 & 0.2173 & 0.1388 & 0.0972 \\ 
  $\hat{\beta}$ & BIAS & -0.0032 & -0.0033 &  0.0040 &  0.2492 &  0.2452 &  0.2355 &  0.0085 &  0.0030 & -0.0024 \\ 
  ~ & SE & 0.1592 & 0.0984 & 0.0775 & 0.2044 & 0.1264 & 0.0872 & 0.2171 & 0.1388 & 0.0972 \\ 
   \hline
\end{tabular}
\caption{RMSE, BIAS and Standard Error (SE) of OLS/CF versus semiparametric.} 
\label{tabols3}
\end{table}

Results are reported in Table \ref{tabols3}. For all estimators, we report the RMSE, the bias and the standard error (SE). The OLS estimator is inconsistent in this setting, and the semiparametric estimator obviously performs better. Moreover, the RMSE of the latter improves also over the control function estimator with a valid continuous instrument. It is interesting to notice that, while the two estimators behave similarly in terms of bias, our estimator leads to a substantial improvement in terms of standard error.

\begin{figure}[!h]
\includegraphics[scale=0.6]{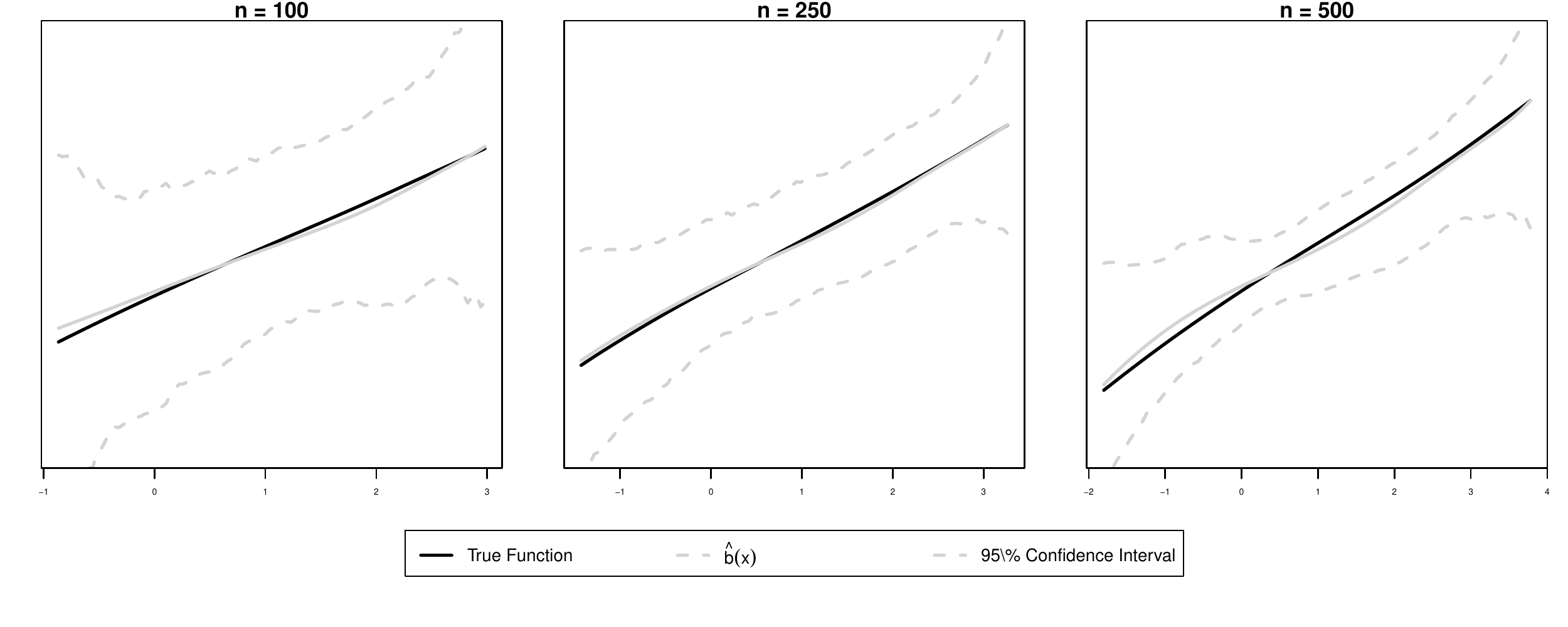}
\caption{Estimator of $b(x)$ for each sample size}
\label{fig:fct_end4}
\end{figure}

We also draw below (see Figure \ref{fig:fct_end4}) the average and the simulated $95\%$ confidence bands for the nonparametric estimator of the function $b(x) = E \left[ \varepsilon_i \vert X_i = x\right]$.

\section{Empirical Application} \label{sec:empiricalapp}

We apply our theoretical results to investigate the relationship between malaria eradication effort and economic outcomes.

Malaria has killed about $430,000$ people in 2015, especially in Sub-Saharan Africa\footnote{Source WHO, \url{https://www.who.int/features/factfiles/malaria/en/}.}. Repeated bouts of malaria can substantially reduce overall children health, and impair their labor productivity as adults. Empirical evidence suggests that countries with malaria tend to be much poorer than countries without malaria. It is however difficult to understand whether malaria is a cause of poverty or viceversa. \citet{bleakley2010} and \citet{cutler2010} both use organized effort to eradicate malaria as a natural experiment to assess its impact on economic outcomes in the US and South America (Brazil, Mexico and Colombia) and India, respectively. First the US in the 1920s and then Latin American countries and India in the 1950s launched major, and successful, eradication campaigns. Both of these papers use a difference-in-difference approach: they compare trends in adult income by birth cohort in regions which did or did not see major reductions in malaria because of the eradication campaigns. 

We focus here on the Colombian data from \citet{bleakley2010}\footnote{Data can be obtained directly from the website of the American Economic Association. See \url{https://www.aeaweb.org/articles?id=10.1257/app.2.2.1}.}. The reason to focus on Colombia is that the sample size is substantially larger (more than $500$ observations) than for other countries, and thus it would reduce concerns over the reliability of our semiparametric procedure.
In this framework, treatment is a continuous variable: regions with higher levels of pre-eradication malaria would potentially benefit more from the campaign. While regions that are Malaria-free would not benefit from the program and would be useful to measure any trend in the chosen economic outcome.
Similarly, Colombians born after 1957 were fully exposed to the eradication campaign. They did not suffer from chronic malaria in their early childhood and consequently they do not have gaps in their education because of the illness. The cohort of people born in 1940 was already in its adulthood by the time the campaign began, and thus serve as a comparison group. 
Individual data are aggregated by Colombian \textit{municipios} (a unit of measure similar to US county), which is our observational unit.\footnote{We refer interested readers to the original paper for a detailed description of the data.}
Among others, \citet{bleakley2010} considers the following specification
\begin{equation}\label{eq:bleakley}
\Delta Y_i = \alpha + X_i \beta  + Z_i \gamma + e_i, 
\end{equation}
where $\Delta Y_i$ is the difference in the economic outcome for the cohort of people born after and before the eradication campaign. For Colombia, \citet{bleakley2010} considers three potential outcomes: literacy, years of schooling and industrial income score. For simplicity, we only consider literacy as potential outcome. The independent variable $X_i$ is the incidence of malaria in the \textit{municipio} before the campaign was started. This is measured by two indices of malaria ecology. The first index (referred to as \textit{Mellinger} in the text) is computed using information on climate and local vectorial capacity \citep{mellinger2004}. The second index (referred to as \textit{Poveda} in the text) measures malaria ecology based on climatic factors \citep{poveda2000}. Finally, $Z_i$ is a vector of regional controls. We only consider the simplest specification in which a group of regional dummies is taken as $Z_i$. Summary statistics for these variables (excluding the regional dummies) are reported in Table \ref{sumstat} below. 
\begin{table}[ht]
\centering
\begin{tabular}{lcccc}
  \hline
 & Mean & St.Dev & Min & Max \\ 
  \hline
$\Delta$ Literacy & 0.02 & 0.12 & -0.30 & 0.76 \\ 
  Mellinger & 0.24 & 0.34 & 0.00 & 1.13 \\ 
  Poveda & 0.46 & 0.42 & 0.00 & 1.01 \\ 
   \hline
\end{tabular}
\caption{Summary statistics} 
\label{sumstat}
\end{table}

In this model, $\alpha$ captures the trend in outcome across \textit{municipios}, and $\beta$ is the average effect of the pre-treatment malaria incidence on the outcome. \citet{bleakley2010} estimates the model in \eqref{eq:bleakley} by OLS and he also constructs an IV estimator to control for the attenuation bias that could be caused by measurement error in $X_i$. Among these IV specifications, he uses the excluded index as instrumental variable for the included index. His results are fairly robust, leading to a positive estimator of $\beta$ irrespective of the outcome used: \textit{municipios} with higher pre-treatment malaria ecology have benefited more from the eradication campaign.

Let us now consider the following extension of the model in \eqref{eq:bleakley}
\begin{equation}\label{eq:bleakley_crc}
\Delta Y_i = \alpha + X_i \beta_i  + Z_{1i} \gamma_i + e_i, 
\end{equation}
where $E \left[ e_i \vert X_i,Z_{1i} \right] = 0$, so that we ignore the measurement error, and we allow the treatment effect to be heterogeneous across \textit{municipios}. One could also argue that, if the technology for eradication is homogeneous, \textit{municipios} with higher level of malaria ecology would require a longer campaign to obtain the same improvement in output. This implies that $\beta_i$ may be correlated with $X_i$, and that this correlation could be negative. 

For illustrative purposes, let
\[
\beta_i = \tilde\beta_0 + X_i \tilde\beta_1 + V_i, 
\]
where $\tilde\beta_0 = \beta - E(X_i) \tilde\beta_1$, with $\beta = E(\beta_i)$, and we take $E \left[ V_i \vert X_i \right] = 0$. We conjecture that $\tilde\beta_1 \leq 0$ in this specification. When plugging this into equation \eqref{eq:bleakley_crc}, and ignoring for simplicity the additional controls, $Z_{1i}$, we obtain

\begin{equation}\label{eq:bleakley_crc2}
\Delta Y_i = \alpha + X_i \tilde\beta_0 +  X^2_i \tilde\beta_1 + V_i X_i + e_i. 
\end{equation}

When one omits the quadratic term from the regression, we have
\begin{align*}
\frac{Cov(X_i,Y_i)}{Var(X_i)} =& \tilde\beta_0  + \tilde\beta_1 \frac{Cov(X_i,X^2_i)}{Var(X_i)}\\
=& \beta - \tilde\beta_1 \left( E(X_i) - \frac{Cov(X_i,X^2_i)}{Var(X_i)} \right).
\end{align*}

Obviously, if $\tilde\beta_1 = 0$, the OLS estimator would be unbiased. However, if $\tilde\beta_1<0$, and 
\[
E(X_i) - \frac{Cov(X_i,X^2_i)}{Var(X_i)} < 0,
\]
as we can infer from the available data, then the bias of the OLS estimator would be negative. This implies that the OLS estimator would consistently underestimate the average treatment effect if there is a negative correlation between the returns to treatment and the pre-treatment level. 

A similar reasoning can be also extended to an IV estimator, when the instrument, $Z_{2i}$, is such that $Cov(Z_{2i},X_i) >0$, and $E(V_i \vert X_i, Z_{2i})= 0$. Notice that the relevance condition as stated is satisfied if one uses the excluded index as instrument. 

We therefore implement our semiparametric procedure to estimate the model in \eqref{eq:bleakley_crc}. Exogenous controls can be accommodated as described in Section \ref{subsec:exo}.

\subsection{Empirical results}

We estimate the following semiparametric model
\[
\Delta Y_{i}= \alpha +  X_{i} \beta + Z_{1i} \gamma + X_{i} b(X_i) + Z_{1i} c(X_i) + e_i + U_i, 
\]
where both $b(X_i)$ and $c(X_i)$ are centered functions of $X$; and 
\[
U_{i} = X_i \left( \varepsilon_i - b(X_i)\right) + Z_{1i} \left( \zeta_i - c(X_i)\right),
\]
which satisfies $E \left[ U_i \vert  X_i,Z_{1i}\right] = 0$, under the additional condition $\lbrace \varepsilon_i,\zeta_i \rbrace \upmodels Z_{1i} \Vert X_i$. Notice that $p=1$, so that in this section too, we have that $b \equiv b^\ast$, and use the notation concurrently. 

The order of polynomials in $X$ for both estimators of $b$ and $c$ is chosen by cross-validation, along the lines of our simulation study. 

We first discuss the results for the parametric part of our model. We do not report coefficients for the control variables $Z$, that are available upon request from the authors.
\begin{table}[ht]
\centering
\begin{tabular}{lcccc}
  \hline ~ & \multicolumn{2}{c}{Mellinger} & \multicolumn{2}{c}{Poveda} \\ \hline
~ & SNP & OLS & SNP & OLS \\ 
  \hline
$\hat{\alpha}$ & -0.0069 & -0.0028 & -0.0195 & -0.0022 \\ 
   & (0.0099) & (0.0089) & (0.0099) & (0.0088) \\ 
   \hline
$\hat{\beta}$ & 0.3077 & 0.0709 & 0.1840 & 0.0354 \\ 
   & (0.0855) & (0.0162) & (0.0412) & (0.0124) \\ 
   \hline
\end{tabular}
\caption{OLS and Semiparametric Estimators (dependent variable Literacy)} 
\label{estres}
\end{table}

Table \ref{estres} reports the results for the estimation of the parameter $\alpha$ and $\beta$ in our model, both with OLS and our semiparametric estimator (compare the former with Table 3, Panel A, p. 20 in \citep*{bleakley2010}). The estimator of the intercept is negative, although not significantly different from zero when we use the Mellinger index. The estimator of $\beta$ is instead positive and significant. The standard errors are computed using our asymptotic results. Compared to the original result of \citet{bleakley2010}, we find a much larger average effect of pre-levels of malaria ecology on literacy.

\begin{figure}[!h]
\includegraphics[scale=0.6]{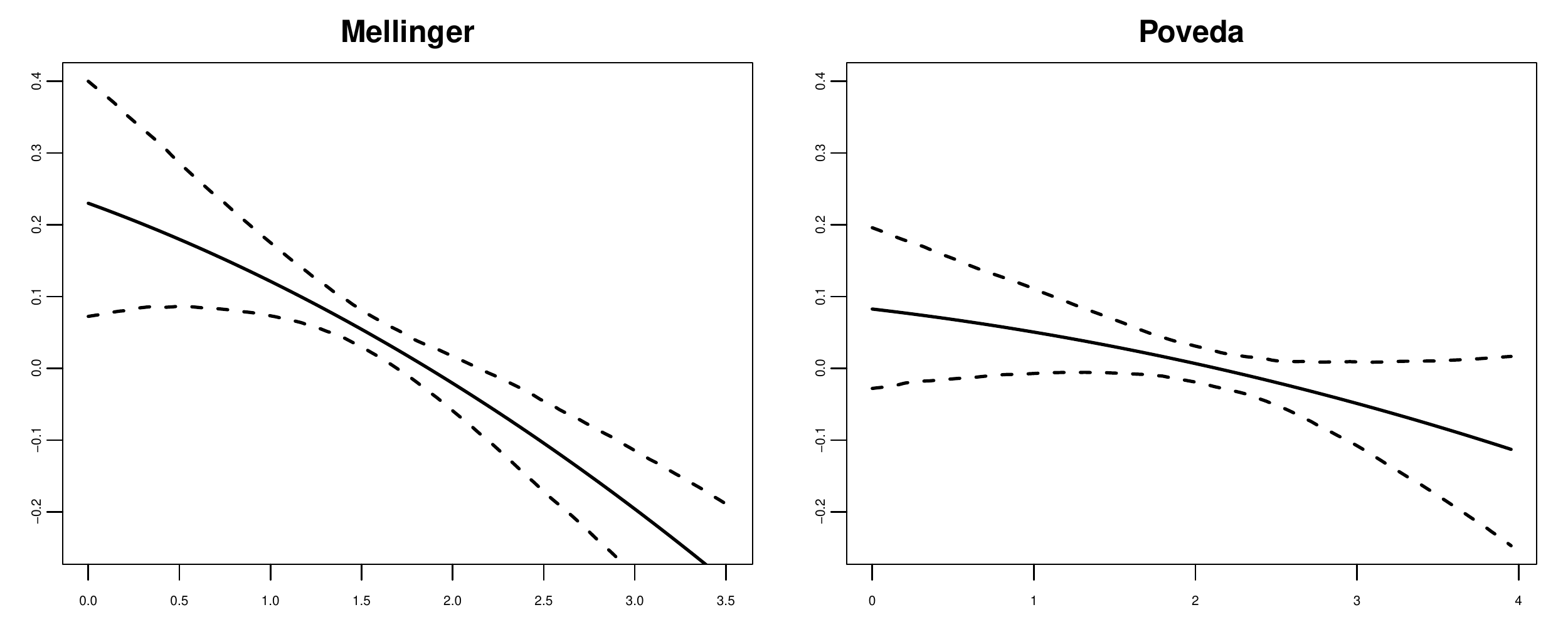}
\caption{Estimator of the conditional mean $\hat{b}(x)$.}
\label{fig:fct_mal}
\end{figure}

In Figure \ref{fig:fct_mal}, we also report the estimator of the conditional expectation of $\varepsilon$ given $X$ for the two indices used. $95\%$ confidence intervals are obtained using wild bootstrap. We have not attempted a formal proof to show that wild bootstrap is valid. However, our framework fits standard assumptions for semiparametric models, and we do not see any potential cause of inconsistency of bootstrap in this context. Notice that both estimators of the function $b(X)$ are negatively sloped, which accords well with our illustration above. Usually, the effect of the eradication campaign would be lower for \textit{municipios} with higher levels of malaria ecology. For the Mellinger index this result is significant, while for the Poveda index, we cannot exclude that the function is flat around $0$, as confidence intervals are substantially wider. 

\section{Conclusions}

In this paper, we study a simple random coefficient model in which the slope coefficients may be correlated with some continuous independent regressors. We show that it is possible to identify and estimate this model without resorting to instrumental variable. The proposed semi-parametric estimator allows one to recover both the Average Partial Effects, and the conditional expectation of the unobserved heterogeneity with respect to these independent regressors. When the dimension of the covariates is greater than one, this is possible under further restrictions on the functional form of the conditional expectation functions. We argue that our estimator could be a valid alternative to existing methods, especially when instrumental variables are not readily available or fail to satisfy exclusion restrictions beyond reasonable doubts. 

\bibliography{../rcm}
\bibliographystyle{apalike}


\newpage

\setcounter{section}{0}
\renewcommand{\thesection}{\Alph{section}}

\section{Appendix}

\subsection{Proof of Propositions \ref{prop:identification} and \ref{prop:identification1}}

We first want to show that, the restrictions imposed by Assumption \ref{ass:identification1} are enough to identify the parameters of the model. That is
\[
W_i \delta + X_i \sum_{j = 1}^p b^\ast_j (X_{j i}) \overset{a.s}{=} 0,
\]
implies that the parameters of the model are almost surely equal to $0$. 

First of all, notice that Assumption 2.1(i) restricts the class of functions to be infinitely differentiable at $0$. Assuming $E(1/X_j^\kappa)$ exists, for $\kappa > 0$, this assumption excludes all functions of the type
\[
b^\ast(x_j) = b_j \left[ \frac{1}{x_j^\kappa} - E\left(\frac{1}{X_j^\kappa} \right) \right],
\]
with $b_j$, a constant, and for all $j= 1,\dots,p$. Therefore, whenever we make claim below of the type $b_j^\ast = 0$, they have to be read as ``there does not exist a function $b_j^\ast \in \mathcal{M}$ that satisfies the restriction, except for the trivial function".

All (in)equalities below are to be intended almost surely, although we shall not make this explicit. (In)equalities for vectors are to be intended component-wise. 

First of all, we notice that, because $b^\ast(x)\in \mathcal{M}$ are centered functions and because of Assumption \ref{ass:identification1}(v), the columns of $W_i$ and $X_i \sum_{j = 1}^p b^\ast_j (X_{j i})$ are not linearly dependent.

We have that
\begin{align*}
E \left[ X^\prime_i X_i \vert X_{j i} = x_j \right] b^\ast_j (X_{j i}) +& E \left[ X^\prime_i X_i \sum_{l = 1,l\neq j}^p b^\ast_l (X_{l i}) \vert X_{j i} = x_j \right] + E \left[ X^\prime_i W_i \vert X_{j i} = x_j \right] \delta  =0 \\
E \left[ W^\prime_i W_i\right] \delta +& E \left[ W^\prime_i X_i \sum_{j = 1}^p b^\ast_j (X_{j i}) \right] = 0.
\end{align*}

Because of Assumption \ref{ass:identification1}(iii)-(iv), the matrices $E \left[ W^\prime_i W_i\right]$ and $E \left[ X^\prime_i X_i \vert X_{j i} = x_j \right]$, for all $j = 1,\dots,p$ are invertible, and we thus can explicitly characterize a solution to this system of equations. We denote this solution $\lbrace b_1^\ast,\delta_1 \rbrace$. 

Now suppose there is another solution $\lbrace b_2^\ast, \delta_2 \rbrace$, such that
\[
W_i \delta_2 + X_i \sum_{j = 1}^p b^\ast_{2j} (X_{j i}) = 0. 
\]

Then, for $j \in 1,\dots,p$, we must have
\begin{align*}
b^\ast_{1j} (x_j) - b^\ast_{2j} (x_j) +& E \left[ X^\prime_i X_i \vert X_{j i} = x_j \right]^{-1} \left( E \left[ X^\prime_i X_i \sum_{l = 1,l\neq j}^p \left( b^\ast_{1l} (X_{l i}) - b^\ast_{2l} (X_{l i})\right) \vert X_{j i} = x_j \right] \right.\\
+& \left. E \left[ X^\prime_i W_i \vert X_{j i} = x_j \right] \left( \delta_1 - \delta_2 \right) \right) = 0\\
\delta_1 - \delta_2 +& E \left[ W^\prime_i W_i\right]^{-1} E \left[ W^\prime_i X_i \sum_{j = 1}^p \left( b^\ast_{1j} (X_{j i}) - b^\ast_{2j} (X_{j i})\right) \right] = 0.
\end{align*}

If $b^\ast_{1l} = b^\ast_{2l} $, for all $l =1, \dots, p$, $l\neq j$, and $\delta_1 = \delta_2$, then we have that  $b^\ast_{1j}=  b^\ast_{2j}$, so that the trivial solution satisfies this system of equations. 

Now suppose that for some $l$, $b^\ast_{1l} \neq b^\ast_{2l}$. Then, we must have that $E \left[ W^\prime_i X_i \left( b^\ast_{1l} (X_{l i}) - b^\ast_{2l} (X_{l i})\right) \right] \neq 0$. 

This assertion follows from the fact that the singular values of the matrix $E \left[ W^\prime_i X_i \vert x_l \right]$ are greater than or equal to the eigenvalues of the matrix $E \left[ X^\prime_i X_i \vert x_l \right]$, and thus bounded away from zero by Assumption \ref{ass:identification1}(iv). 
This finally implies that $\delta_1 \neq \delta_2$, and therefore, for all $j$, $b^\ast_{1j} \neq b^\ast_{2j}$. This concludes the proof. 

\subsection{Preamble}

We follow closely the notations and the proof in \citet{ahmad2005}, \citet{belloni2014} and \citet{chenchristensen2013}. We first introduce some notations. 

Let $G^{\oplus}_{\psi,j}$ be $pK \times pK$ matrix formed by the direct sum of $\lbrace G_{\psi,j^\prime }, j^\prime = 1,\dots,p\rbrace$. Similarly, let $\mathbf{G}^{\oplus}_{\psi}$, the $p^2 K \times p^2 K$ matrix formed by the direct sum of $\lbrace G^{\oplus}_{\psi,j}, j = 1,\dots,p\rbrace$. Notice that, by the properties of block diagonal matrices and Assumption \ref{assest2}(i), the smallest eigenvalues of $G^{\oplus}_{\psi,j}$ and $\mathbf{G}^{\oplus}_{\psi}$ are also bounded away from $0$, for all $K > 0$. It also immediately follows that
\begin{align*}
\sup_{\xi \in [0,1]}& \Vert  \tilde{\Psi}^\oplus (\xi) \mathbf{G}^{\oplus,-1/2}_{\psi} \Vert_{\ell^2}  \leq \max_{j} \sup_{\xi \in [0,1]} \Vert  \tilde{\Psi}^\oplus_{j}(\xi) G^{\oplus,-1/2}_{\psi,j} \Vert_{\ell^2} \\
\leq &\max_{j,j^\prime}\sup_{\xi \in [0,1]} \Vert  \tilde{\psi}_{j^\prime j K}(\xi) G^{-1/2}_{\psi,j^\prime j}\Vert_{\ell^2} \leq \zeta_K.
\end{align*}

We let $\hat{\SM}_W = \mathbf{W}^\prime_n \SM_n/n$, and
\begin{align*}
\PM^o =& \mathbf{G}^{\oplus,-1/2}_{\psi}\PM \mathbf{G}^{\oplus,-1/2}_{\psi} \\
\hat{\PM}^o_n =& \mathbf{G}^{\oplus,-1/2}_{\psi}\hat{\PM}_n \mathbf{G}^{\oplus,-1/2}_{\psi}
\end{align*}
and
\begin{align*}
\SM^o_W =& E \left[ W^\prime_i S(X_i) \right] \mathbf{G}^{\oplus,-1/2}_{\psi},\\
\hat{\SM}^o_W =& \hat{\SM}_W \mathbf{G}^{\oplus,-1/2}_{\psi}.
\end{align*}

Finally, $G^{o}_{\psi,j} = I_{pK}$, the identity matrix of dimension $pK$.

\subsection{Proof of Theorem \ref{thconvasnorm}}
Using the notations previously defined, we can rewrite
\begin{align*}
\hat{\delta} =& \left(\frac{\mathbf{W}_n^\prime \mathbf{W}_n}{n}-  \hat{\SM}_W \hat{\PM}_n^{-}  \hat{\SM}^\prime_W \right)^{-1}\left( \frac{\mathbf{W}_n^\prime \mathbf{Y}_n}{n}  -\hat{\SM}_W \hat{\PM}_n^{-} \frac{\SM_n^\prime \mathbf{Y}_n}{n} \right),\\
\hat{\pi} =& \hat{\PM}_n^{-} \left( \SM^\prime_n \mathbf{Y}_n/n - \hat{\SM}^\prime_W \hat{\delta}\right).
\end{align*}

Finally, we let
\[
\mathbf{B}^\ast_n = \begin{bmatrix} X_1 \sum_{j = 1}^p b^\ast_j(X_{j1}) \\ \vdots \\  X_n \sum_{j = 1}^p b^\ast_j(X_{jn}) \end{bmatrix}.
\]

From equation \eqref{eq:deltahat}, we have 
\begin{align*}
\sqrt{n} \left( \hat{\delta} - \delta \right) =& \sqrt{n} \left( \frac{\mathbf{W}^\prime_n \mathbf{W}_n}{n}  - \hat{\SM}_W \hat{\PM}_n^{-1} \hat{\SM}^\prime_W \right)^{-1}\left( \mathbf{W}^\prime_n  - \hat{\SM}_W \hat{\PM}_n^{-1} \SM^\prime_n  \right) \left( \mathbf{B}^\ast_n + \mathbf{U}_n\right)/n \\
=& \left( \frac{\mathbf{W}^\prime_n \mathbf{W}_n}{n}  - \hat{\SM}_W \hat{\PM}_n^{-1} \hat{\SM}^\prime_W \right)^{-1} \sqrt{n} \left( \mathbf{W}^\prime_n  - \hat{\SM}_W \hat{\PM}_n^{-1} \SM^\prime_n  \right)  \mathbf{B}^\ast_n/n \\
+& \left( \frac{\mathbf{W}^\prime_n \mathbf{W}_n}{n}  - \hat{\SM}_W \hat{\PM}_n^{-1} \hat{\SM}^\prime_W \right)^{-1} \sqrt{n} \left( \mathbf{W}^\prime_n  - \hat{\SM}_W \hat{\PM}_n^{-1} \SM^\prime_n  \right)  \mathbf{U}_n/n.
\end{align*}

From Lemma \ref{lem:app2b}, we have that 
\[
\frac{\mathbf{W}^\prime_n \mathbf{W}_n}{n}  - \hat{\SM}_W \hat{\PM}_n^{-1} \hat{\SM}^\prime_W  = \Phi + o_P(1).
\]
From Lemma \ref{lem:app3}, the bias term is such that
\[
\sqrt{n} \left( \mathbf{W}^\prime_n  - \hat{\SM}_W \hat{\PM}_n^{-1} \SM^\prime_n  \right)  \mathbf{B}^\ast_n/n = O_P\left( \sqrt{n} s_K \right) = o_P(1).
\]
To conclude the proof, we apply Lemma \ref{lem:app4} to the variance term, and the result of the Theorem follows. 

\subsection{Proof of Theorem \ref{thconvproof}}

We only prove the second part of the Theorem (uniform convergence). The first part (convergence in mean-squared error) follows immediately from the results in \citet{newey1997}, and \citet{ahmad2005}, along with the new developments in \citet{belloni2014} and \citet{chenchristensen2013}, who loosen the requirements on smoothing parameters for the approximation of the design matrix using recent results in random matrix theory \citep{tropp2015}.

We let
\begin{align*}
\tilde{b}^{\ast,K}(\xi) =& \tilde{\mathbf{\Psi}}^\oplus(\xi) \PM^{-1} E \left[ S (X_i)^\prime \sum_{j = 1}^p b^\ast_j(X_{ji}) \right] = \tilde{\mathbf{\Psi}}^\oplus(\xi) \pi_0.
\end{align*}

We can decompose
\begin{align*}
& \hat{b}^{\ast,K}(\xi) - b^\ast(\xi) = \hat{b}^{\ast,K}(\xi) - \tilde{b}^{\ast,K}(\xi) + \tilde{b}^{\ast,K}(\xi)  - b^\ast(\xi)  \\
=& \tilde{\mathbf{\Psi}}^\oplus(\xi) \hat{\PM}_n^{-} \left( \SM^\prime_n \mathbf{Y}_n/n - \hat{\SM}^\prime_W \hat{\delta}\right) - \tilde{\mathbf{\Psi}}^\oplus(\xi) \pi_0 \\
+& \tilde{\mathbf{\Psi}}^\oplus(\xi) \pi_0 - b^\ast(\xi)\\
=& \tilde{\mathbf{\Psi}}^\oplus(\xi) \hat{\PM}_n^{-} \SM^\prime_n \mathbf{U}_n/n \tag{$T_1(\xi)$} \\
+& \tilde{\mathbf{\Psi}}^\oplus(\xi) \hat{\PM}_n^{-} \SM^\prime_n\left( \mathbf{B}^\ast_n - \SM_n \pi_0  \right)/n \tag{$T_2(\xi)$}  \\
-& \tilde{\mathbf{\Psi}}^\oplus(\xi) \hat{\PM}_n^{-} \hat{\SM}^\prime_W \left( \hat{\delta} - \delta\right) \tag{$T_3(\xi)$}\\
+& \tilde{\mathbf{\Psi}}^\oplus(\xi) \pi_0 - b^\ast(\xi) \tag{$T_4(\xi)$}.
\end{align*}

Directly from Assumption \ref{assest2}(iii), we obtain
\[
\sup_{\xi \in [0,1]} \vert T_4(\xi) \vert = O_P\left( N_K s_K \right).
\]

We now turn to the remaining terms. We write
\begin{align*}
T_{2}(\xi) =&  \tilde{\mathbf{\Psi}}^\oplus(\xi) \hat{\PM}_n^{-} \SM^\prime_n\left( \mathbf{B}^\ast_n - \SM_n \pi_0  \right)/n  \\
=& \tilde{\mathbf{\Psi}}^\oplus(\xi) \mathbf{G}^{\oplus,-1/2}_{\psi} \left( \hat{\PM}_n^{o-} - \PM^{o-1} \right) \mathbf{G}^{\oplus,-1/2}_{\psi} \SM^\prime_n\left( \mathbf{B}^\ast_n - \SM_n \pi_0  \right)/n \\
+& \tilde{\mathbf{\Psi}}^\oplus(\xi) \mathbf{G}^{\oplus,-1/2}_{\psi}\PM^{o-1} \mathbf{G}^{\oplus,-1/2}_{\psi} \SM^\prime_n\left( \mathbf{B}^\ast_n - \SM_n \pi_0  \right)/n \\
=& T_{21}(\xi) + T_{22}(\xi).
\end{align*}

First, we notice that 
\[
\Vert \PM^{o-1} \mathbf{G}^{\oplus,-1/2}_{\psi} \SM^\prime_n\left( \mathbf{B}^\ast_n - \SM_n \pi_0  \right)/n \Vert_{\ell^2} = O_P \left( \sqrt{\frac{K}{n}} N_K s_K\right),
\]
from a direct bound and Assumptions \ref{assest2}(ii) and \ref{assest3}(iii). Therefore
\begin{align*}
\sup_{\xi \in [0,1]} \vert T_{22}(\xi)  \vert  \leq&   \sup_{\xi \in [0,1]}  \Vert  \tilde{\mathbf{\Psi}}^\oplus(\xi) \mathbf{G}^{\oplus,-1/2}_{\psi} \Vert_{\ell^2} \Vert  \PM^{o-1} \mathbf{G}^{\oplus,-1/2}_{\psi} \SM^\prime_n\left( \mathbf{B}^\ast_n - \SM_n \pi_0  \right)/n \Vert_{\ell^2}  \\
\leq& O_P\left( \zeta_K \sqrt{\frac{K}{n}} N_K s_K \right) = O_P(N_K s_K),
\end{align*}
for $\zeta_K \sqrt{K/n} = O_P(1)$. From the same bound and Lemma \ref{lem:app1a}, we have that
\begin{align*}
\sup_{\xi\in [0,1]} \vert T_{21}(\xi) \vert  = o_P(N_K s_K),
\end{align*}

The term in $T_3(\xi)$ can be treated equally, with $\hat{\delta} - \delta= O_P(n^{-1/2})$, so that
\[
\sup_{\xi\in [0,1]} \vert T_{3}(\xi)  \vert   = O_P(\zeta_K \sqrt{\frac{K}{n}}n^{-1/2} ) = O_P(n^{-1/2}).
\]
Finally, we have
\begin{align*}
T_{1}(x) =&  \tilde{\mathbf{\Psi}}^\oplus(\xi) \hat{\PM}_n^{-} \SM^\prime_n \mathbf{U}_n/n    \\
=& \tilde{\mathbf{\Psi}}^\oplus(\xi) \left( \hat{\PM}_n^{-} - \PM^{-1} \right) \SM^\prime_n \mathbf{U}_n/n  \\
+& \tilde{\mathbf{\Psi}}^\oplus(\xi) \PM^{-1} \SM^\prime_n \mathbf{U}_n/n \\
=& T_{11}(\xi) + T_{12}(\xi).
\end{align*}
Directly from Assumptions \ref{assest2}(iii) and \ref{assest3}(i), Lemma \ref{lem:app2}, and the result above, we have that
\[
\sup_{\xi\in [0,1]} \vert T_{11}(\xi)  \vert   = O_P\left( \zeta_K \sqrt{\frac{\log K}{n}}\right).
\]
Similarly, using Lemmas \ref{lem:app1a} and \ref{lem:app1b}, and the results of Proposition 6.1 and Theorem 6.1 in \citet{belloni2014} 
\[
\sup_{\xi\in [0,1]} \vert T_{12}(\xi)  \vert   = O_P\left( \zeta_K \sqrt{\frac{\log K}{n}}\right).
\]
The result of the Theorem follows.


\subsection{Additional Lemmas}
\begin{lemma} \label{lem:app2b}
Let Assumptions \ref{assest1}-\ref{assest2} hold, with $\sqrt{n} s_K = o_P(1)$. Then
\[
\frac{\mathbf{W}^\prime_n \mathbf{W}_n}{n}  - \hat{\SM}_W \hat{\PM}_n^{-} \hat{\SM}^\prime_W \xrightarrow{p} E \left[ \left( W_i - E_\mathcal{V} \left[ W_i\right] \right)^\prime \left( W_i - E_\mathcal{V} \left[ W_i\right] \right) \right] 
\]
\begin{proof}
Notice that directly from the application of the LLN and Assumption \ref{assest1}(i), we have that
\[
\frac{\mathbf{W}^\prime_n \mathbf{W}_n}{n} \xrightarrow{p} E \left[ W_i^\prime W_i \right].
\]
We now deal with the second term. Notice that we can rewrite
\[
\hat{\SM}^o_W \hat{\PM}^{o,-} \hat{\SM}^{o\prime}_W - \SM^o_W \PM^{o,-1} \SM^{o\prime}_W + \SM^o_W \PM^{o,-1} \SM^{o\prime}_W.
\]
The first term can be bound using Lemmas \ref{lem:app1a} and \ref{lem:app1b} with
\begin{align*}
\Vert \hat{\SM}^o_W \hat{\PM}^{o,-} \hat{\SM}^{o\prime}_W - \SM^o_W \PM^{o,-1} \SM^{o\prime}_W \Vert_{\ell^2}  = O_P \left( \zeta_K \sqrt{\frac{\log K}{n}}\right) = o_P(1).
\end{align*}
Finally, notice that $E_\mathcal{V} \left[ W_i \right]$ can be written using a generalized Fourier decomposition on the vector of basis functions $\tilde{\psi}_K$, properly normalized. Therefore, by Assumptions \ref{ass:smoothness}, and \ref{assest2}(iv),
\[
\Vert S \PM^{o,-1} \SM^{o\prime}_W - E_\mathcal{V} \left[ W_i \right] \Vert_2 = O_P(s_K).
\]
Thus
\begin{align*}
& \Vert E \left[ \SM^{o}_W  \PM^{o,-1} S^\prime (X_i) S(X_i) \PM^{o,-1} \SM^{o\prime}_W  - E_\mathcal{V} \left[ W_i \right]^\prime E_\mathcal{V} \left[ W_i \right] \right] \Vert_2 \\
\leq& E\Vert S \PM^{o,-1} \SM^{o\prime}_W - E_\mathcal{V} \left[ W_i \right] \Vert^2_2 = O(s^2_K) = o(1). 
\end{align*}
This concludes the proof. 
\end{proof}

\end{lemma}

\begin{lemma} \label{lem:app3}
Let Assumptions \ref{assest1}-\ref{assest2} hold. Then
\[
\sqrt{n} \left( \mathbf{W}^\prime_n  - \hat{\SM}_W \hat{\PM}_n^{-1} \SM^\prime_n  \right)  \mathbf{B}^\ast_n/n = O_P \left( \sqrt{n} s_K\right) = o_p(1)
\]
\begin{proof}
Notice that
\begin{align*}
& \Vert \left( \mathbf{W}^\prime_n  - \hat{\SM}_W \hat{\PM}_n^{-1} \SM^\prime_n  \right)  \mathbf{B}^\ast_n \Vert_{\ell^2} \\
\leq &\sup_{x \in [0,1]^p} \Vert w \Vert_{\ell^2} \sup_{x \in [0,1]^p} \Vert x \Vert_{\ell^2}  \sum_{j = 1}^p \Vert b^\ast_j(\xi) - \tilde{\Psi}_j^{\oplus}(\xi) \hat{\PM}_n^{-} \SM^\prime_n \mathbf{B}^\ast_n/n \Vert_{\ell^2} \\
& = O_P(s_K),
\end{align*}
directly by Theorem \ref{thconvproof}. The result follows. 
\end{proof}
\end{lemma}

\begin{lemma} \label{lem:app4}
\[
n^{-1/2} \left( \mathbf{W}^\prime_n  - \hat{\SM}_W \hat{\PM}_n^{-1} \SM^\prime_n  \right)  \mathbf{U}_n  \overset{d}{\longrightarrow} N\left( 0 , \Omega \right)
\]
\begin{proof}
\begin{align*}
n^{-1/2} &\left( \mathbf{W}^\prime_n  - \hat{\SM}_W \hat{\PM}_n^{-1} \SM^\prime_n  \right)  \mathbf{U}_n  = n^{-1/2} \sum_{i = 1}^n W^\prime_i \left( U_i - S(X_i) \hat{\PM}^{-} \frac{1}{n}\sum_{i^\prime = 1}^n S(X_{i^\prime})^\prime U_{i^\prime}\right).
\end{align*}
Because of Lemmas \ref{lem:app1a}, \ref{lem:app2} and the LLN, we have that
\begin{align*}
\hat{\PM} =& \PM + o_P(1),\\
\frac{1}{n} \sum_{i = 1}^n W^\prime_i S(X_i) \mathbf{G}^{\oplus,-1/2}_{\psi} =& \SM^o_W + o_P(1).
\end{align*}
Therefore, we can write 
\begin{align*}
n^{-1/2} & \sum_{i = 1}^n W^\prime_i \left( U_i - S(X_i) \hat{\PM}^{-} \frac{1}{n}\sum_{i^\prime = 1}^n S(X_{i^\prime})^\prime U_{i^\prime}\right) \\
=& n^{-1/2}  \sum_{i = 1}^n W^\prime_i U_i - \SM^o_W \PM^{o,-1} n^{-1/2} \sum_{i^\prime = 1}^n \mathbf{G}^{\oplus,-1/2}_{\psi} S(X_{i^\prime})^\prime U_{i^\prime}+ o_P(1).
\end{align*}
Furthermore, from a direct application of CLT, Assumption \ref{assest1}(ii), and from the proof of Lemma \ref{lem:app2}, we obtain
\[
\SM^o_W \PM^{o,-1} n^{-1/2} \sum_{i = 1}^n  \mathbf{G}^{\oplus,-1/2}_{\psi} S(X_{i^\prime})^\prime U_{i^\prime} \overset{d}{\longrightarrow} N\left( 0, E\left[ \sigma^2 (X_i) E_\mathcal{V}\left[ W_i \right]^\prime E_\mathcal{V}\left[ W_i \right] \right] \right).
\]
Similarly, 
\[
n^{-1/2} \sum_{i = 1}^n W^\prime_i U_i \overset{d}{\longrightarrow} N\left( 0, E\left[ \sigma^2 (X_i) W_i^\prime W_i \right] \right),
\]
and the result of the Lemma directly follows from the definition of $\Omega$ in equation \eqref{eq:omegadef}.
\end{proof}
\end{lemma}

\begin{lemma} \label{lem:app1a}
Let Assumptions \ref{assest1}-\ref{assest2} hold, and let $\hat{G}_{\psi,j}$, be the sample counterpart of $G_{\psi,j}$. Then it follows that
\begin{align*}
\Vert \hat{G}^o_{\psi,j} - I_{K} \Vert_{\ell^2} =& O_P\left( \zeta_K \sqrt{\frac{\log K}{n}}\right),
\end{align*}
as $n,K \rightarrow \infty$, with $\zeta_K \sqrt{\log K/n} \rightarrow 0$. 
\begin{proof}
See \citet{belloni2014} and \citet{chenchristensen2013}, who prove this result using the Bernstein inequality for random matrices given in Lemma \ref{lem:add1}. 
\end{proof}
\end{lemma}

\begin{lemma} \label{lem:app1b}
Let Assumptions \ref{assest1}-\ref{assest2} hold. Then
\begin{align*}
\Vert \hat{\PM}_n^o - \PM^o \Vert_{\ell^2} =& O_P\left( \zeta_K \sqrt{\frac{\log K}{n}}\right),\\
\Vert \hat{\SM}^o_W - \SM^o_W \Vert_{\ell^2} =& O_P\left( \zeta_K \sqrt{\frac{\log K}{n}}\right).
\end{align*}
\begin{proof}

To prove the first bound let
\[
\sum_{i = 1}^n n^{-1} \mathbf{G}^{\oplus,-1/2}_{\psi} \left\lbrace S(X_i)^\prime S(X_i) -  E \left[  S(X_i)^\prime S(X_i)  \right]\right\rbrace \mathbf{G}^{\oplus,-1/2}_{\psi} = \sum_{i = 1}^n \Upsilon_i,
\]
with $\Vert \Upsilon_i \Vert_{\ell^2} \leq 2n^{-1} \zeta^2_K$, and 
\begin{align*}
\left\Vert \sum_{i = 1}^n E \left[ \Upsilon^\prime_i \Upsilon_i \right] \right\Vert_{\ell^2}  =& \left\Vert \sum_{i = 1}^n E \left[   \Upsilon_i \Upsilon^\prime_i\right] \right\Vert_{\ell^2} \\
\leq & n^{-1} \Vert  \mathbf{G}^{\oplus,-1/2}_{\psi} E \left[ \tilde{\mathbf{\Psi}}^{\oplus,\prime}(X_i) X^\prime_i X_i \tilde{\mathbf{\Psi}}^\oplus(X_i)  \mathbf{G}^{\oplus,-1}_{\psi} \tilde{\mathbf{\Psi}}^{\oplus,\prime}(X_i) X_i^\prime X_i \tilde{\mathbf{\Psi}}^\oplus(X_i) \right]\mathbf{G}^{\oplus,-1/2}_{\psi} \Vert_{\ell^2}  \\
\leq& n^{-1} \zeta^2_K \Vert  \mathbf{G}^{\oplus,-1/2}_{\psi} E \left[ \tilde{\mathbf{\Psi}}^{\oplus,\prime}(X_i) \tilde{\mathbf{\Psi}}^\oplus(X_i) \right]  \mathbf{G}^{\oplus,-1/2}_{\psi}  \Vert_{\ell^2}\\ \leq& n^{-1} \zeta^2_K,
\end{align*}
where the second last line follows from Assumption \ref{assest1}(i), which implies $\sup_{x \in [0,1]^p} \Vert x \Vert^2_{\ell^2} <\infty$.

Similarly, for the second bound, we redefine
\[
\sum_{i = 1}^n n^{-1} \left\lbrace  W^\prime_i S(X_i) -  E \left[  W^\prime_i S(X_i) \right]\right\rbrace \mathbf{G}^{\oplus,-1/2}_{\psi} = \sum_{i = 1}^n \Upsilon_i,
\]
with $\Vert \Upsilon_i \Vert_{\ell^2} \leq 2n^{-1} \zeta_K$.
We thus have
\begin{align*}
\left\Vert \sum_{i = 1}^n E \left[  \Upsilon_i \Upsilon^\prime_i\right] \right\Vert_{\ell^2} \leq & n^{-1} \Vert E \left[ W_i^\prime X_i \tilde{\mathbf{\Psi}}^{\oplus}(X_i) \mathbf{G}^{\oplus,-1}_{\psi} \tilde{\mathbf{\Psi}}^{\oplus,\prime}(X_i) X_i^\prime W_i \right] \Vert_{\ell^2}  \\
\leq& n^{-1} \zeta^2_K,
\end{align*}
where the second last line is again implied by Assumption \ref{assest1}(i) with
\[
\sup_{x} \langle w,x \rangle_{\ell^2} < \sup_{x} \Vert x \Vert_{\ell^2} \sup_{x} \Vert w \Vert_{\ell^2} < \infty.
\]
Finally,
\begin{align*}
\left\Vert \sum_{i = 1}^n E \left[ \Upsilon^\prime_i\Upsilon_i \right] \right\Vert_{\ell^2} \leq & n^{-1} \Vert \mathbf{G}^{\oplus,-1/2}_{\psi}  E \left[  \tilde{\mathbf{\Psi}}^{\oplus,\prime}(X_i) X_i^\prime W_i W_i^\prime X_i \tilde{\mathbf{\Psi}}^{\oplus}(X_i) \right] \mathbf{G}^{\oplus,-1/2}_{\psi} \Vert_{\ell^2}  \\
\leq& n^{-1} \Vert \mathbf{G}^{\oplus,-1/2}_{\psi} E \left[ \tilde{\mathbf{\Psi}}^{\oplus,\prime}(X_i) \tilde{\mathbf{\Psi}}^{\oplus}(X_i) \right] \mathbf{G}^{\oplus,-1/2}_{\psi}  \Vert_{\ell^2} \leq n^{-1}.
\end{align*}
The second bound follows from $\max \lbrace \zeta^2_K,1 \rbrace = \zeta^2_K$. 
\end{proof}
\end{lemma}

\begin{lemma} \label{lem:app2}
Suppose that Assumptions \ref{ass:smoothness}-\ref{assest3} hold. Then
\[
\Vert G^{-1/2}_{\Psi} \mathbf{S}_n^\prime \mathbf{U}_n/n \Vert_{\ell^2} = O_P(\sqrt{K/n}).
\]
\begin{proof}
See \citet{newey1997}.
\end{proof}
\end{lemma}

\begin{lemma}[Bernstein inequality for random matrices \citep{tropp2015}] \label{lem:add1}
Let $S_1,\dots,S_n$ be independent, centered random matrices with common dimension $d_1 \times d_2$, and assume that each one is uniformly bounded
\[
ES_k = 0 \text{ and } \Vert S_k \Vert_{\ell^2} \leq L \quad \text{ for each } k=1,\dots,n.
\]
Introduce the sum
\[
Z = \sum_{k =1}^n S_k,
\]
and let $v(Z)$ denote the matrix variance statistic of the sum:
\begin{align*}
v(Z) =& \max \left\lbrace \Vert E \left( Z Z^\prime \right) \Vert_{\ell^2},\Vert E \left( Z^\prime Z  \right) \Vert_{\ell^2}\right\rbrace \\
=& \max \left\lbrace \left\Vert \sum_{k = 1}^n E \left( S_k S_k^\prime \right) \right\Vert_{\ell^2},\left\Vert \sum_{k = 1}^n E \left( S_k^\prime S_k  \right) \right\Vert_{\ell^2}\right\rbrace.
\end{align*}
Then
\[
\mathbb{P} \lbrace \Vert Z  \Vert_{\ell^2} \geq t \rbrace \leq (d_1 + d_2 ) \exp \left\lbrace \frac{-t^2/2}{v(Z) + Lt/3}\right\rbrace \quad \text{for all } t \geq 0,
\]
and
\[
E \Vert Z \Vert_{\ell^2} \leq \sqrt{2 v(Z)  \log\left( d_1 + d_2 \right)} + \frac{1}{3}L\log\left( d_1 + d_2 \right).
\]
\end{lemma}
\end{document}